\documentclass[conference]{IEEEtran}
\IEEEoverridecommandlockouts
\usepackage{cite}
\usepackage{amsmath,amssymb,amsfonts}
\usepackage{algorithm}[H]
\usepackage{algorithmic}
\usepackage{algorithm}
\usepackage[utf8]{inputenc}
\usepackage{graphicx}
\usepackage{textcomp}
\usepackage{xcolor}

\usepackage{tcolorbox}
\usepackage{enumitem}
\usepackage{nccmath}
\usepackage{lipsum}
\usepackage{booktabs}
\usepackage{multicol}
\usepackage{multirow} 
\usepackage{makecell}
\usepackage{pifont}
\usepackage{verbatim}
\usepackage{subfigure} 

\usepackage{amsthm}

\newtheorem{prop}{Proposition}

\setlist[itemize]{leftmargin=*}
\setlist[enumerate]{leftmargin=*}
\usepackage{fancyhdr}

\usepackage{hyperref}
\hypersetup{
    colorlinks=true, 
    linktoc=all,     
    linkcolor=blue,  
}
\def\BibTeX{{\rm B\kern-.05em{\sc i\kern-.025em b}\kern-.08em
    T\kern-.1667em\lower.7ex\hbox{E}\kern-.125emX}}
\begin{document}
\newcommand{\SWITCH}[1]{\STATE \textbf{switch} (#1) \textbf{do}}
\newcommand{\ENDSWITCH}{\STATE \textbf{end switch}}
\newcommand{\CASE}[1]{\STATE \textbf{case} #1\textbf{:} \begin{ALC@g}}
\newcommand{\ENDCASE}{\end{ALC@g}}
\newcommand{\CASELINE}[1]{\STATE \textbf{case} #1\textbf{:} }
\newcommand{\DEFAULT}{\STATE \textbf{default:} \begin{ALC@g}}
\newcommand{\ENDDEFAULT}{\end{ALC@g}}
\newcommand{\DEFAULTLINE}[1]{\STATE \textbf{default:} }

\newcommand{\PARALLEL}[1]{\STATE \textbf{parallel for} #1 \textbf{do}}
\newcommand{\ENDPARALLEL}{\STATE \textbf{end parallel}}

\title{\textit{``And Then There Were None''}: Cracking White-box DNN Watermarks via Invariant Neuron Transforms}

\author{\IEEEauthorblockN{Yifan Yan, Xudong Pan, Yining Wang, Mi Zhang, Min Yang}
\IEEEauthorblockA{School of Computer Science \\
\textit{Fudan University, China}\\
\{yanyf20,xdpan18,18307130267,mi\_zhang,m\_yang\}@fudan.edu.cn}}

\maketitle
\pagestyle{plain}
\lhead{}
\chead{}
\rhead{}
\lfoot{}
\cfoot{}
\rfoot{\thepage}

\begin{abstract}
Recently, how to protect the Intellectual Property (IP) of deep neural networks (DNN) becomes a major concern for the AI industry. To combat potential model piracy, recent works explore various watermarking strategies to embed secret identity messages into the prediction behaviors or the internals (e.g., weights and neuron activation) of the target model. Sacrificing less functionality and involving more knowledge about the target model, the latter branch of watermarking schemes (i.e., white-box model watermarking) is claimed to be accurate, credible and secure against most known watermark removal attacks, with emerging research efforts and applications in the industry.

In this paper, we present the first effective removal attack which cracks almost all the existing white-box watermarking schemes with provably no performance overhead and no required prior knowledge. By analyzing these IP protection mechanisms at the granularity of neurons, we for the first time discover their common dependence on a set of fragile features of a local neuron group, all of which can be arbitrarily tampered by our proposed chain of invariant neuron transforms. On $9$ state-of-the-art white-box watermarking schemes and a broad set of industry-level DNN architectures, our attack for the first time reduces the embedded identity message in the protected models to be almost random. Meanwhile, unlike known removal attacks, our attack requires no prior knowledge on the training data distribution or the adopted watermark algorithms, and leaves model functionality intact.  
\end{abstract}




\section{Introduction}
In the past decade, the ever-increasing size of deep neural networks (DNNs) incurs rapid growth of model training costs\cite{he2016resnet, radford2015dcgan, szegedy2016inception, zagoruyko2016wrn, xu2015caption,devlin2018bert, strubell2019energy}, which incentivizes model piracy attacks \cite{oh2019reverseNN, orekondy2019knockoff, wang2018stealhyper, shokri2017membership,salem2018mlleaks}. To trace illegal model copies, \textit{DNN watermarking} emerges as a promising approach and arouses wide research interests \cite{uchida2017embedding, wang2021riga,liu2021greedyresiduals, fan2021deepip, zhang2020passportaware, ong2021iprgan,chen2021lottery, lim2022ipcaption,darvish2019deepsigns,adi2018turning,zhang2018blackwatermark, szyller2021dawn, jia2021entangled}. In a general model watermarking process, a secret identity message is first embedded into \textit{the target model} during the training stage (i.e., \textit{watermark embedding}). Later, the ownership is verified if the same or a similar watermark is detected from \textit{a suspect model} (i.e., \textit{watermark verification}).

Depending on how the suspect model is accessed during verification, existing watermarks are categorized into \textit{black-box} and \textit{white-box} schemes. Generally, a black-box watermark is based on the model's prediction behavior on a subset of data inputs which are priorly chosen by the watermarking scheme\cite {adi2018turning,zhang2018blackwatermark, szyller2021dawn, jia2021entangled}, while a white-box watermark is based on the model internals, including the model parameters \cite{uchida2017embedding,  wang2021riga,liu2021greedyresiduals, chen2021lottery,fan2021deepip, zhang2020passportaware, ong2021iprgan} and the neuron activation \cite{darvish2019deepsigns, lim2022ipcaption}. Sacrificing less functionality and involving more knowledge about the suspect model, white-box model watermarking is widely considered more comprehensive compared with the black-box counterpart \cite{wang2021riga,liu2021greedyresiduals, fan2021deepip, jia2021entangled}, with emerging applications and research efforts from many industry leaders (e.g., Microsoft \cite{zhang2020passportaware, darvish2019deepsigns, chen2019deepattest}).
\begin{table}[t]
  \caption{Our attack strategy successfully cracks $9$ mainstream white-box DNN watermarks.}
    \centering
  \scalebox{0.75}{
    \begin{tabular}{clccccc}
    \toprule
    \multirow{2}[4]{*}{\textbf{Year}} & \multirow{2}[4]{*}{\textbf{Method}} & \multicolumn{4}{c}{\textbf{Existing Attacks}} & \multicolumn{1}{c}{\multirow{2}[4]{*}{\textbf{Ours}}} \\
\cmidrule{3-6}          &       & \multicolumn{1}{l}{Pruning} & \multicolumn{1}{l}{Fine-tuning} & \multicolumn{1}{l}{Overwriting} & \multicolumn{1}{l}{Abiguity} &  \\

    \midrule
    2017 & Uchida et al.\cite{uchida2017embedding} & \ding{55}     & \ding{55}      & \ding{51}   & \ding{51}    & \ding{51}  \\
    \midrule
    2019 & DeepSigns\cite{darvish2019deepsigns} & \ding{55}     & \ding{55}      & \ding{55}      & \ding{51}   & \ding{51}\\
    \midrule
    2020 & Passport-Aware\cite{zhang2020passportaware} & \ding{55}     & \ding{55}     & \ding{55}    & \ding{55}      & \ding{51}  \\
    \midrule
\multirow{5}[4]{*}{2021} & DeepIPR\cite{fan2021deepip} & \ding{55}     & \ding{55}      & \ding{55}   & \ding{55}     & \ding{51}  \\
    \cmidrule{2-7}
    & RIGA\cite{wang2021riga}  & \ding{55}     & \ding{55}      & \ding{55}      & \ding{55}   & \ding{51}  \\
    \cmidrule{2-7}
     & Greedy Residuals\cite{liu2021greedyresiduals} & \ding{55}     & \ding{55}      & \ding{55}     & \ding{55}    & \ding{51}  \\
     \cmidrule{2-7}
     & IPR-GAN\cite{ong2021iprgan} & \ding{55}     & \ding{55}     & \ding{55}  & \ding{55}       & \ding{51} \\
     \cmidrule{2-7}
     & Lottery Verification\cite{chen2021lottery} & \ding{55}     & \ding{55}    & \ding{55}     & \ding{55}      & \ding{51}  \\
    \midrule
    2022 & IPR-IC\cite{lim2022ipcaption} & \ding{55}     & \ding{55}      & \ding{55}     & \ding{55}   & \ding{51}  \\
    \bottomrule
    \end{tabular}}%
  \label{tab:intro_table}%
\end{table}%


In a typical threat scenario, an attacker with a stolen DNN would modify the model parameters to frustrate the success of watermark verification \cite{see2016compression, li2016filterpruning, hinton2015distilling, wang2019overwrite, yang2019distillremoval, shafieinejad2021robustofbackdoorbased, chen2021refit, aiken2021laundering, guo2021ftnotenough, wang2019neuralcleanse}. As a major trade-off for ownership obfuscation, the attacker inevitably encounters slight degradation on the normal model utility \cite{adi2018turning}. In the current literature, consensus is reached that recent white-box watermarks have strong resilience against such \textit{watermark removal attacks} \cite{zhang2020passportaware,fan2021deepip,liu2021greedyresiduals,ong2021iprgan,chen2021lottery}. On the one hand, relying on the internals of the suspect model, the embedded identity messages in white-box watermarking are much strongly connected with the model performance, compared with the external prediction behaviors on a special subset of input-output pairs used in black-box watermarking. 
Therefore, the removal process usually incurs larger decrease in model performance for white-box watermarks than for black-box ones. For example, machine unlearning techniques help remove a subset of prediction rules from a suspect model with almost no utility loss, while similar attacks are not applicable to white-box watermarks thanks to the intrinsic relations between the parameters and the identity messages \cite{ shafieinejad2021robustofbackdoorbased, chen2021refit, aiken2021laundering, guo2021ftnotenough}. On the other hand, many known attack attempts via conventional post-processing techniques (e.g., fine-tuning and pruning  \cite{hinton2015distilling, see2016compression, li2016filterpruning}) also report empirical success on black-box model watermarks, but inevitably perturb the model parameters at an unacceptable scale to fully remove a white-box model watermark, resulting in a DNN with poor performance\cite{uchida2017embedding, wang2021riga,liu2021greedyresiduals, chen2021lottery, fan2021deepip, zhang2020passportaware, ong2021iprgan,darvish2019deepsigns, lim2022ipcaption}. 

\noindent\textbf{Our Work.}  With increasingly more white-box watermarking techniques exploring the usage of the strong correlation to the significant parameters or selected stable activation maps to claim even stronger resilience than previous approaches\cite{ wang2021riga,liu2021greedyresiduals, chen2021lottery,fan2021deepip,lim2022ipcaption,zhang2020passportaware,ong2021iprgan}, the full transparency of the suspect model further strengthens the impression that white-box model watermarks are accurate, credible and secure, with almost no overhead on model performance. However, we find it is a false sense of security. Our work strikingly shows, most existing white-box DNN watermarks can \textit{be fully and blindly removed from a watermarked model with no loss on the model utility}.



For the first time, we reveal exploiting \textit{invariant neuron transforms} \cite{geometryfeedforward, neyshabur2015pathsgd, ganju2018property, bui2020relunetworksequivalence} is a simple yet effective attack strategy against the mainstream white-box model watermarking schemes, most of which heavily rely on a set of fragile features, including the orders, the magnitudes and the signs of a local group of neurons for watermark embedding. By applying a chain of invariant neuron transforms, we are able to arbitrarily manipulate the message embedded in the local feature while provably preserving the normal functionality of the model. As Table \ref{tab:intro_table} shows, we successfully crack $9$ representative white-box DNN watermarks, which claim high robustness against some or all known removal attacks, reducing the embedded watermark to be almost random. In comparison, almost no known existing attacks can simultaneously crack even $1/3$ of them. Besides, our attack leaves the model utility provably intact after watermark removal. Moreover, as a \textit{blind} attack, our approach requires no knowledge on the training data distribution, the adopted watermark embedding process, or even the existence of a watermark.




Technically, the core building blocks of our attack framework are three types of invariant neuron transforms, namely \textit{LayerShuffle}, \textit{NeuronScale}, and \textit{SignFlip}, which manipulate three types of local features, i.e., \textit{orders, magnitudes and signs}, of specific weights or neuron activation. We find these local features of model parameters are heavily utilized in existing white-box model watermarking schemes, but are highly fragile under the corresponding transforms. To crack the existing white-box watermark schemes, we construct a chain of invariant transforms on the model parameters which reduces the watermark embedded in one set of local features to randomness, while compensating for the malicious perturbations on watermark-related parameters by calibrating other related local features. This yields a functionally equivalent model to the original watermarked one, yet without the watermark.

Specifically, we first rearrange the local order through \textit{LayerShuffle}, which applies an arbitrary permutation to the neurons within each selected hidden layer to reshuffle the pertinent model parameters. Then, we alter the local magnitude by \textit{NeuronScale}, which scales up/down the incoming edges to each randomly chosen hidden neuron with a positive real number, and scales down/up the outgoing edges with the same ratio correspondingly. Finally, we perturb the local sign with \textit{SignFlip}, which targets at flipping the signs of model weights arbitrarily sampled from the watermarked model, while adjusting the signs of related parameters in adjacent layers (i.e., a linear/convolutional layer or a normalization layer\cite{ioffe2015batch, wu2018gn, ba2016layer, ulyanov2016instance}).
As these three operations are not mutually exclusive and complementary to each other, we integrate them into a comprehensive attack framework to crack popular white-box watermarking verification \textit{blindly}, i.e., without knowing the adopted watermark scheme and even the existence of a watermark. Also, unlike previous removal attacks based on fine-tuning or retraining \cite{adi2018turning, chen2021refit, guo2021ftnotenough}, our attack has no requirement on training data throughout the attack by design.


To validate the effectiveness of our attack scheme, we present a systematic study on 9 existing white-box watermarking methods, each of which claims high robustness against previous removal attacks in their original evaluation\cite{uchida2017embedding, wang2021riga,liu2021greedyresiduals, chen2021lottery, fan2021deepip, zhang2020passportaware, ong2021iprgan,darvish2019deepsigns, lim2022ipcaption}. According to our case-by-case analysis, most of them share the common vulnerability against invariant neuron transforms. Consequently, our attack achieves almost perfect watermark removal on all the evaluated white-box watermarking methods and, as is certified by our analytical proofs, incurs no influence on the model functionality. We strongly hope future works on white-box model watermarking should avoid relying on local features of neurons and derive more robust and resilient features, which should be at least invariant under the three types of invariance we exploit.


\noindent\textbf{Our Contributions.} In summary, we mainly make the following contributions:
\begin{itemize}
\item We reveal almost all the existing white-box model watermarking schemes share the common vulnerability of relying on a set of fragile features derived from local neuron groups (\S\ref{sec:rethink}), which we find severely suffers from being arbitrarily tampered by invariant neuron transforms.
\item We propose the first watermark removal attack which applies a chain of random invariant neuron transforms to remove the white-box watermark from a protected model, incurs provably no utility loss, and requires no prior knowledge on the watermark embedding and the training processes (\S\ref{sec:attack_framework}). 

\item We validate the success of our attack on a wide group of industry-level DNNs protected by $9$ state-of-the-art white-box watermarking schemes, some of which are presented on top-tier conferences by the industry (e.g., Microsoft \cite{zhang2020passportaware, darvish2019deepsigns, chen2019deepattest}). The extracted identity messages are reduced to almost random after our removal attack, while the normal model utility remains the same (\S\ref{sec:case_study}).
\end{itemize}

\section{Preliminary}
\subsection{Deep Neural Networks} 
In this paper, we focus on the watermarking of deep neural networks (DNNs), a popular family of learning models which are used to fit decision functions $f(\cdot;W)(:=f_W):\mathcal{X}\to\mathcal{Y}$ with layers of learnable parameters $W$. A DNN is composed of layers of computation performed by feed-forward layers (e.g., fully connected or convolutional layers), which maps an input in $\mathcal{X}$ to an output in $\mathcal{Y}$ iteratively by applying predefined computing formulas with the learnable parameters. The intermediate layers extract the hidden representation features sequentially along with a nonlinear activation function (e.g., ReLU) or pooling function (e.g., average pooling). To obtain the optimal DNN model which fits the training data and generalizes to the unknown test data, the parameters are updated iteratively during the training phase by optimizing the loss function with off-the-shelf gradient-based optimizers such as stochastic gradient descent (SGD). 

Given a DNN with $H$ layers, i.e., $f_W = \{f^1, f^2, \hdots,f^H\}$, each layer $f^l$ is composed of a set of $N_l$ neurons ($f^l=(n_1^l, n_2^l,...,n_{N_l}^l)$). We denote the parameters of the $l^{th}$ $(1\leq l\leq H)$ layer as $W^l$,
which can be further written as $\{w^l_{ij}\}_{i=1,j=1}^{N_{l-1}, N_l} \cup \{b^l_{j}\}_{j=1}^{N_l}$. From a neuron-level viewpoint, each element $w^l_{ij}$ is a real-valued weight parameter in fully connect layers or a kernel matrix in convolutional layers, which connects the neurons $n^{l-1}_i$ and $n^l_j$, and $b^l_{j}$ is the bias.


\subsection{Normalization Layer}
\label{sec:normlization layer}
In commercial deep learning models, the normalization layers become an increasingly indispensable design component\cite{he2016resnet, radford2015dcgan, szegedy2016inception, zagoruyko2016wrn, xu2015caption,  devlin2018bert}. Intuitively, a normalization layer is first proposed for accelerating the training process and is later proved to be beneficial for boosting the performance of a DNN model by a significant margin. In the past decade, multiple normalization schemes have been proposed to cater for different requirements, while Batch Normalization (BN) remains one of the most widely-used methods \cite{ioffe2015batch,wu2018gn,ba2016layer,ulyanov2016instance}. To reduce the internal covariate shift, BN conducts normalization along the batch dimension to control the distribution of intermediate features, while other normalization layers, including Layer Normalization (LN), Instance Normalization (IN) and Group Normalization (GN), exploit the channel dimension to pursue more improvement in model performance. 

Specifically, the aforementioned normalization layers, which are practically inserted after convolutional layers or linear layers, first normalize the data features by the running mean $\mu$ and variance $\sigma^2$ of the historical training data, and then conduct a linear transformation using a scale factor $\gamma$ and a bias shift $\beta$, which follows the common formula: $
\hat{x} = \gamma \frac{x - \mu}{\sigma} + \beta,$
where $\gamma$, $\beta$ are learnable parameters. We highlight that BN continuously monitors and updates the $\mu$ and the $\sigma$ during the training process while fixing the values during inference, whereas other normalization layers update statistic values on the fly throughout both training and testing time. As a result, proper normalization layer parameters are critical for reproducing the reported accuracy of the model when deployed. Even slight perturbation to these parameters would cause catastrophic damage to the model functionality by deviating hidden features from the correct distribution. Inspired by this phenomenon, a recent branch of white-box model watermarks also explore to encode the secret information into the parameters of selected normalization layers\cite{ong2021iprgan,fan2021deepip,zhang2020passportaware}.


\subsection{White-box Model Watermarking} \label{sec: white-box wm}
Confronting with the urgent need of IP protection on DNN, a number of model watermarking strategies are developed to enable the legal owner of a DNN to claim the ownership by verifying the existence of a unique watermark embedded in a suspect model. In this paper, we focus on the white-box watermarking schemes, where the model parameters and activation maps are accessible during the verification.

These approaches consist of two stages: \textit{watermark embedding} and \textit{watermark verification}. In the former stage, the owner of a model $f_W$ encodes the secret message $s$ (e.g., a binary string) into the model parameters or intermediate activation maps by adding a regularization term $\mathcal{L}_{EM}$ to the primary learning objective $\mathcal{L}_{Ori}$, i.e.,
$
\mathcal{L} = \mathcal{L}_{Ori} + \lambda \mathcal{L}_{EM},
$
where $\lambda$ is the hyper-parameter related to watermark embedding procedure. For example, $\mathcal{L}_{EM}$ in \cite{uchida2017embedding} is the binary cross entropy loss between the secret bit string $s$ and $\sigma(X\cdot w)$, where $w$ is derived from the parameters of a specified convolutional layer by channel-level averaging and flattening, $X$ is a predefined transformation matrix, and $\sigma$ is sigmoid function. In the latter stage, an equal-length message $s'$ is extracted by a white-box extraction function $\mathcal{E}$ in polynomial time:
\begin{equation}\label{eq:extract wm}
s' = \mathcal{E}(f_W, M, A),
\end{equation}
where $M$ is the mask matrix to select a set of the specific parameters or activation maps directly from the target model $f_W$, and $A$ is a transformation function which finally projects the selected weights or activations to obtain the extracted message $s'$. For example, \cite{uchida2017embedding} extract the message from the watermarked model according to the signs of the relative parameter after projection, i.e., $s' = T_h(X\cdot w) $, where $T_h$ is a hard threshold function which outputs $1$ when the input is positive and $0$ otherwise and $w$ is the weights selected from model $f_W$ by mask matrix $M$.
Therefore, the owner can prove the model ownership once the distance (e.g., Hamming Distance \cite{hamming1950error}) between the extracted message $s'$ and the secret message $s$ is less than a predefined threshold $\epsilon$.

Furthermore, in order to ensure a secure and trustworthy ownership verification, a model watermarking scheme should satisfy the minimum set of requirements as follows:
\begin{itemize}
\item \textbf{Fidelity.} The watermark embedding procedure should have little negative impact on the model's initial utility.

\item \textbf{Reliability.} If the suspect model is the same or a post-processed copy of the watermarked model, the watermark verification procedure should identify the model as a theft model with high confidence; otherwise, the suspect model should be detected as an irrelevant model owned by others to yield minimal false positives.

\item \textbf{Robustness.} The watermark inside the model should be resistant to any removal attack, such as model post-processing techniques (e.g., fine-tuning or pruning). Moreover, it should not be trivial for the adversary to embed a fabricated watermark into the target model via ambiguity attacks, which raises doubts to the ownership verification \cite{fan2021deepip}.
\end{itemize}
\section{Previous Knowledge on Model Watermark Security} 
\subsection{Attack Taxonomy}
Existing watermarking techniques are generally questioned for their practical robustness. A line of works attempt to defeat these watermarking mechanisms and can be divided into two types: \textit{ambiguity attacks} and \textit{removal attacks}. First proposed by Fan et al. \cite{fan2021deepip}, the ambiguity attack aims at forging a counterfeit watermark to cast a doubt on the ownership verification without modifying the original model weights. Instead, the removal attack focuses on discrediting the verification technique by erasing the traces of the legal owner completely. As our work is a removal attack against the mainstream white-box watermarking techniques, we assume the threat model for the adversary who attempts to remove the embedded watermarks in Section \ref{sec:threat_model}, and discuss the limitations of existing removal attacks in Section \ref{sec:limitations}.

\subsection{Threat Model of Model Watermark Removal Attacks}
\label{sec:threat_model}
In our threat model, the adversary has obtained an illegal copy of a watermarked model which allows full access to its model parameters. Such model piracy can be accomplished via either algorithmic attacks \cite{tramer2016StealViaApi, yu2020cloudleak} or system attacks exploiting software/hardware vulnerabilities \cite{jeong2021meltdown,yan2020cache}. To conceal the traces of model infringement, the attacker attempts to invalidate the model ownership verification by removing the existing watermarks. To impose the least assumptions on the adversarial knowledge, we propose a strong removal attack should fulfill the following requirements:

\begin{itemize}
\item \textbf{No Influence on Model Functionality:} The primary goal of the adversary is to obtain the target model utility. As a result, the attack should have almost no impact on the original model performance.

\item \textbf{No Knowledge on Watermarking Process:} The adversary should have no knowledge about the adopted watermarking embedding and extraction algorithms, which are usually exclusively known to the owner until the verification is launched on a suspect model.

\item \textbf{No Access to Original Training Data:} As the training is usually a private asset of the model owner, the access to the original training data or even data from a similar domain, is not always practical in reality. Therefore, our work restricts the attacker from obtaining any knowledge of confidential distribution of the training data, since he/she otherwise may legally train his own model from scratch, which makes the white-box watermark removal attack unnecessary.
\end{itemize}

\subsection{Existing Removal Attacks and Their Limitations}
\label{sec:limitations}
Mainly via perturbing the parameters related to the inside watermark, existing removal attacks to the white-box model watermarking techniques can be divided into three groups: 
\begin{itemize}

\item \textbf{Weight pruning} is an efficient post-processing technique to reduce the storage and accelerate the computation process of DNNs, which sets a proportion of redundant parameters to zero without degrading the original accuracy. However, previous white-box watermark works show the high robustness to pruning-based attacks, while effective parameter pruning need to remove a certain number of connections between the neurons in DNNs, along with an unacceptable loss to the model performance\cite{uchida2017embedding, darvish2019deepsigns}.

\item \textbf{Fine-tuning} is another common watermark removal attack studied in earlier robustness evaluation, which continues the training operation for a few steps without considering the watermarking specific loss functions. This method requires a small set of data from the original training set or following a similar distribution, otherwise the model utility would suffer from a noticeable degradation. Furthermore, a certain amount of computational resources are also expected to support this extra model training \cite{chen2021refit, guo2021ftnotenough}.

\item \textbf{Overwriting} is first proposed in \cite{wang2019overwrite} to show the vulnerability of the classical model watermark algorithm in \cite{uchida2017embedding}. Considering the adversary has full knowledge about the watermarking process, he/she may remove the existing watermark by embedding a new one related to his own identification information. However, the details of watermark schemes, including the embedding algorithm, extraction procedure and the hyper-parameters used for training are always not available in real-world settings. Further, previous research shows that the original watermark can still be extracted for the legal owner to file an infringement lawsuit \cite{wang2021riga, liu2021greedyresiduals}. Moreover, for several advanced white-box watermarking methods, overwriting attacks even fail to encode a new adversarial message into the target model\cite{fan2021deepip, ong2021iprgan, lim2022ipcaption, zhang2020passportaware}.
\end{itemize} 
We do not consider the model extraction attacks (e.g., \cite{Jagielski2020HighAA}) in our work, as this branch of works mainly aim at stealing models from prediction APIs and also require a substantial amount of training cost to distill a well-trained surrogate models from the victim model. In summary, none of existing removal attacks could meet all the requirements in Section \ref{sec:threat_model} simultaneously. Meanwhile, these attacks are proven to be more effective on black-box model watermarking instead, probably because the black-box model watermarking only rely on the prediction results on a specific set of inputs and hence the behavior can be altered without concerns on the otherwise constraints on modifying some internal parameters which may be related with the white-box model watermarking \cite{ shafieinejad2021robustofbackdoorbased, chen2021refit, aiken2021laundering, guo2021ftnotenough, wang2019neuralcleanse}. With more novel white-box watermarking techniques proposed recently claiming even stronger robustness under limited evaluation, white-box model watermarking attracts increasing attention from the academy and industry as more robust and desirable for practical usages. 

\subsection{Our Findings on Model Watermark Security}
On the contrary, our work for the first time discovers and systematizes the innate vulnerabilities in white-box model watermarking, i.e. the local neuron invariance in watermarked models. We present the first attack framework to fully remove the white-box model watermark while incurring certifiably no influence on the model functionality, blindly applicable to most existing white-box model watermarking schemes, and requiring no prior knowledge on the details of watermark schemes or the distribution of training data. Our finding strongly indicates that evaluating robustness against limited attacks in earlier work is insufficient.
First, our removal attack has no effect on the model's functionality, whereas common removal attacks against white-box watermarking by perturbing the model parameters (e.g. fine-tuning or pruning) always cause a measurable utility loss, which is undesirable in some mission-critical applications. Second, as we validate in Section \ref{sec:case_study}, our removal attack is effective to all the known white-box watermarking schemes as a post-processing method, while the adversary is unaware of the details of employed watermarking technique. Third, the results of our extensive work demonstrate that the data and computing power for fine-tuning are unnecessary to invalidate the white-box model watermarking verification, where a underlying commonality overlooked by the most researchers results the root vulnerability of these white-box watermark schemes.


\section{Rethinking the Robustness of White-box Model Watermark}
\label{sec:rethink}
\subsection{Intrinsic Locality of White-box Watermarks}
\label{sec:existing_wmk}
By systematically analyzing the mechanisms of existing white-box watermarking schemes, we discover their common vulnerability of relying on the features of a local neuron group and their connecting weights, which are however fragile under malicious manipulation. We name this newly discovered vulnerability as the \textit{intrinsic locality} of white-box watermarks, which refers to three types of local features, i.e., the signs, the magnitudes and the orders of the connecting weights. 
Considering an arbitrary group of model parameters, i.e., $\mathcal{A}=\{w_i\}^N_{i=1}$, the local features are correspondingly extracted via $Sign$, $Abs$ and $Ind$, which are element-wise functions applied to each weight $w_i$. Specifically, $Sign$ and $Abs$ respectively extract the sign and the absolute value of a real number, while $Ind$ returns the subscript index of $w_i$ in the parameter group (i.e., $ind(w_i) = i$). Then we adopt these functions to extract three kinds of representations from a group of model parameters, i.e., $\{Sign(w_i)\}^N_{i = 1}$, $\{Abs(w_i)\}^N_{i = 1}$ and $\{Ind(w_i)\}^N_{i = 1}$, which we call the \textit{local sign, local magnitude} and \textit{local order} of the connecting weight $w_i$. In the following, we show most of existing white-box watermarking techniques heavily rely on these types of intrinsic locality.

Based on whether specific parameters or intermediate activation maps are selected for watermark embedding and verification, existing white-box watermarking schemes are broadly divided into two groups: \textit{weight-based} and \textit{activation-based}. According to (\ref{eq:extract wm}), these two types of watermarking schemes extract the embedded information from the target model $f_W$ through a mask matrix $M$ and a transformation matrix (or function) $A$, which is then used to compute the similarity score with the secret watermark offered by the legitimate owner. 
In general, the extraction procedure first derives the watermark by applying  a set of watermark-related weights (or activation values) of the target model based on 0–1 values of the mask matrix, and extracts the local features of each element. Formally, it writes $\Phi(M \cdot f_W) = M\cdot \Phi(f_W)$ for $\Phi \in \{Sign, Abs, Ind\}$. Then, a transformation matrix $A$ is designed to extract the final identity message from the derived local feature to determine the model ownership. 
For example, to calculate the identity message, almost all the transformation procedures in existing white-box watermarking algorithms begin with averaging along certain (but not all) dimensions, the result of which can be easily influenced when the signs, magnitudes and orders of the watermark-related parameters (or activation) are modified. To prove it, we consider an average operation along the second dimension denoted as $\textit{Avg}(X) \in \mathbb {R}^{d_1}$, where $X \in \mathbb{R}^{d_1\times d_2}$ is the related weight matrix. As a result, the sign, magnitude and order in the first dimension of each parameter in $X$ will directly influence the final averaged matrix.
Similarly, other possible operations in the transformation matrix $A$ including linear transforms and the \textit{Sign} function are all impacted by the local features of the specific parameters (or activation). Section \ref{sec:case_study} presents a more detailed case-by-case analysis.

In recent literature, some research works start to study \textit{passport-based} white-box watermarking schemes, which in general prove the model ownership by preserving the high performance of the target model after correctly replacing the normalization layers with the private passport layers during the verification procedure. Correspondingly, the passport-based watermarks also rely on the local features of the weights right before the replaced layers, as the activation maps should remain unchanged to match the following scale and bias parameters of the private passport layers (Section \ref{sec:case_study}). 

In summary, almost all the white-box watermarking verification procedures are by design based on the strong correlation between the local features of a certain group of neurons (model parameters or selected activation maps) and the embedded identity message. However, we for the first time discover it is able to construct a sequence of equivalent transformations to the target model, which applies malicious perturbations to the watermark-related parameters to crack the watermark while calibrating other unrelated model intrinsic locality to exactly preserve the prediction behavior of the model. Such a novel attack technique provably breaks most existing white-box watermarking schemes, severely casting serious doubt on our previous knowledge on their robustness. 



\subsection{Invariant Neuron Transforms}
\label{sec:dnn_invariance}

\begin{figure}[t]
\begin{center}
\includegraphics[width=0.45\textwidth]{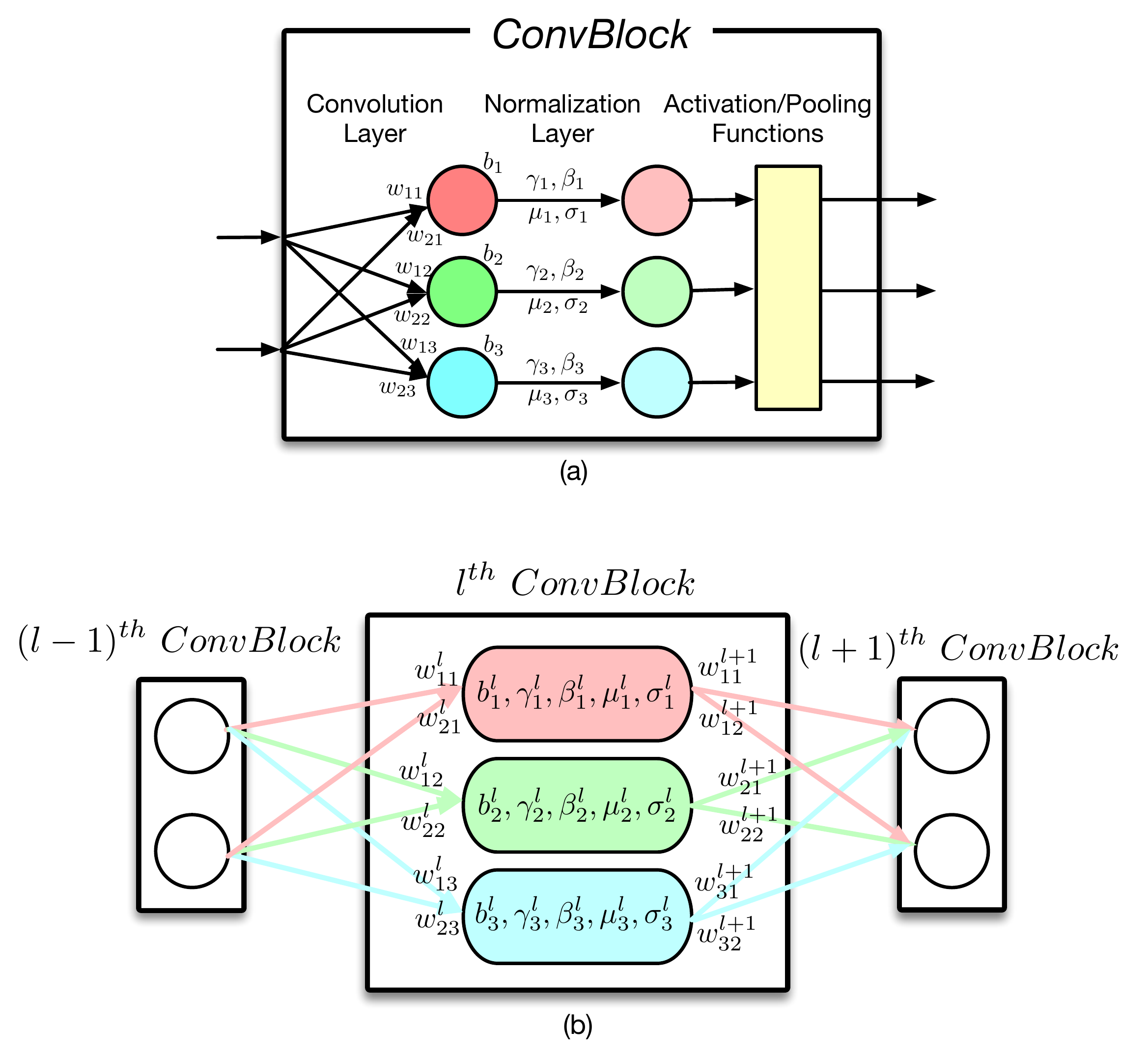}
\caption{A \textit{ConvBlock} (a) is a sequence of intermediate layers and (b) is connected with other ConvBlock modules with incoming and outgoing weights.}
\label{fig:convblock}
\end{center}
\vspace{-0.3in}
\end{figure}
Below, we introduce three types of invariant neuron transforms on DNN which are the building blocks to our proposed attack framework. As a motivating example, we demonstrate the invariant properties of DNNs with a standard \textit{ConvBlock} as Fig.\ref{fig:convblock} shows, which subsequently consists of a convolutional layer, a normalization layer, an activation function and a pooling function. Following the typical designs in many off-the-shelf DNN architectures, we consider a \textit{ConvBlock} with BN layer and ReLU activation. Given the input data $x \in \mathbb{R}^{N_{l-1} \times D_1 \times  D_2}$ from the preceding layer, the output $f^l$ of the $l^{th}$ \textit{ConvBlock} layer can be formulated as follows:

\begin{equation}\label{eq:ConvBlock}
f^l(x) = pool \cdot \sigma_{ReLU} (\gamma^l \frac{W^l\odot x + b^l - \mu^l}{\sigma^l} + \beta^l),
\end{equation}
where $\odot$ denotes the convolution operator, \textit{pool} is the pooling function (max pooling unless otherwise specified), $\sigma_{ReLU}$ is the activation function, $\gamma^l, \beta^l \in \mathbb{R}^{N_l}$ are learnable parameters for normalization, $\mu^l, \sigma^l \in \mathbb{R}^{N_l}$ are statistic values (fixed in testing for BN), and $W^l \in \mathbb{R}^{N_{l-1}\times  N_l \times  k_1 \times  k_2}, b^l \in\mathbb{R}^{N_l}$ are the weights of original convolutional layer in the $l^{th}$ \textit{ConvBlock} layer when $k_1$ and $k_2$ are the width
and height of each filter. 

Similarly, a \textit{LinearBlock} layer consists of one linear layer, the following normalization layer, and an activation function. As a fully connected layer in DNNs is a simplified form of convolutional layer, we focus on showing the invariant property in \textit{ConvBlock} layer and all the operations or formulations can be adapted to the fully connected layer. Moreover, as the state-of-the-art DNNs are typically a collection of \textit{ConvBlock} layers and \textit{LinearBlock} layers, the following invariance property in the prototypical \textit{ConvBlock} also commonly exists in more complicated DNN models with even other normalization layers or activation functions (\S\ref{sec:discussion}). 



In a specific \textit{ConvBlock} layer, the inner layers with convolution or normalization parameters contain the same number of neurons as the output channels, and the following pure arithmetic functions are neuron-level operations without any parameters. As a result, a \textit{ConvBlock} layer can be regarded as a vector of neurons $\{n_i\}_{i = 1} ^{N_l}$ according to previously defined notation for the neurons in DNNs.
The neuron $n^l_i$ is associated with the parameters in current layer (i.e., $w^l_{\cdot i}, b^l_i, \gamma^l_i, \beta^l_i, \mu^l_i, \sigma^l_i$) and the next layer (i.e., $w^l_{i \cdot}$) directly, which we simply denoted as \textit{incoming weights} $W^{l+1}_{i,in}$ and \textit{outgoing weights} $W^l_{i,out}$, respectively. Also, the activation map $h^l_i$ of the neuron $n^l_i$ with input $x$ can be written as follows:

\begin{equation}\label{eq:ConvBlock_i}
h^l_i(x; W^l_{i,in}) = max(\sigma_{ReLU} (\gamma^l_i \frac{w^l_{\cdot i}\odot x + b^l_i - \mu^l_i}{\sigma^l_i} + \beta^l_i) ),
\end{equation}
where the max pooling function $pool$ in \eqref{eq:ConvBlock} is converted to the max function.
Then we introduce three operations on the incoming weights and outgoing weights of randomly selected neurons to obtain an equivalence of the DNN, which will produce exactly the same result to the target model.

\noindent$\bullet$\textbf{ LayerShuffle.} \label{sec:layershuffle}
\begin{figure}[t]
\begin{center}
\includegraphics[width=0.45\textwidth]{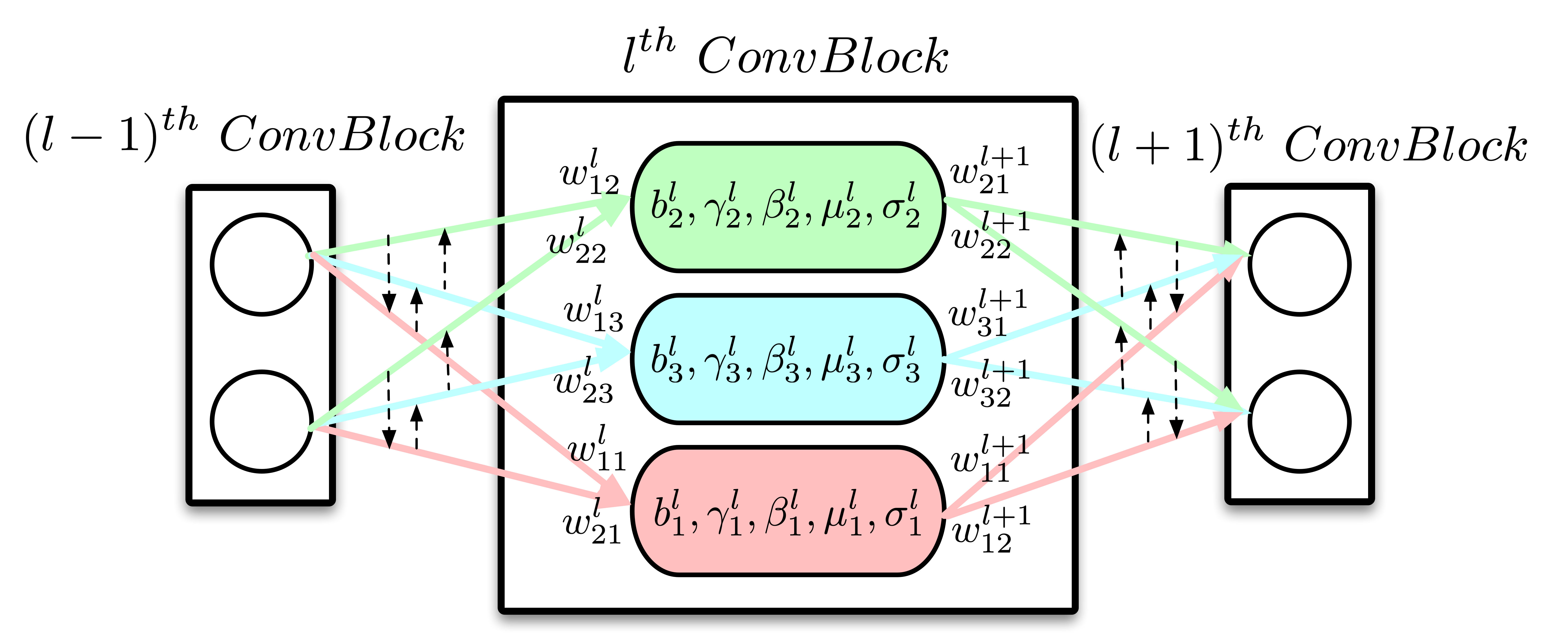}
\caption{An example of LayerShuffle with p=\{2,3,1\}.}
\label{fig:layershuffle}
\end{center}
\end{figure}
In order to define the shuffling equivalence, we first denote a permutation function $p$ on a vector of elements $X = \{x_1, x_2, ... , x_h\}$ as $X \cdot p = \{x_{p(1)}, x_{p(2)}, ... , x_{p(h)}\}$, where $\{{p(i)}\}^h_{i=1}$ is a random permutation of integers from 1 to $h$. Then we can randomly shuffle the nodes in each intermediate layer $f^l$ with the incoming weights and outgoing weights to obtain an equivalent model \cite{ganju2018property}. As shown in Fig.\ref{fig:layershuffle}, we denote the 
LayerShuffle operation on the $l^{th}$ \textit{ConvBlock} layer of model $f$ with a random permutation $p$ ($|p|=N_l$) as $\mathcal {LS}(f^l, p)$ to obtain a new model $f'$. This operation first applies the permutation $p$ on the incoming weights to obtain shuffled activation map as $h^{l}_i{}'=h^{l}_i(x; W^{l}_{p(i),in}) = h^{l}_{p(i)}$, and then shuffle the outgoing weights, i.e., $w^{l+1}_{ij}{}' = w^{l+1}_{p(i)j}$, to produce the same activation map into the original order. We present the rigorous proofs to the following propositions on the invariance of neuron transforms in Appendix \ref{sec:app:proof}.

\begin{prop}
\label{prop: layer_shuffling}
Consider a deep neural network $f$ and a new deep neural network $f'$ obtained from f by a series of LayerShuffle operations $\mathcal {LS}(f^l, p_l)$ to a collection of intermediate layers picked at random. Then, for every input $x$, we have $f(x) = f' (x)$.
\end{prop}
\noindent$\bullet$\textbf{ NeuronScale.} \label{sec:neuronscale}
\begin{figure}[t]
\begin{center}
\includegraphics[width=0.45\textwidth]{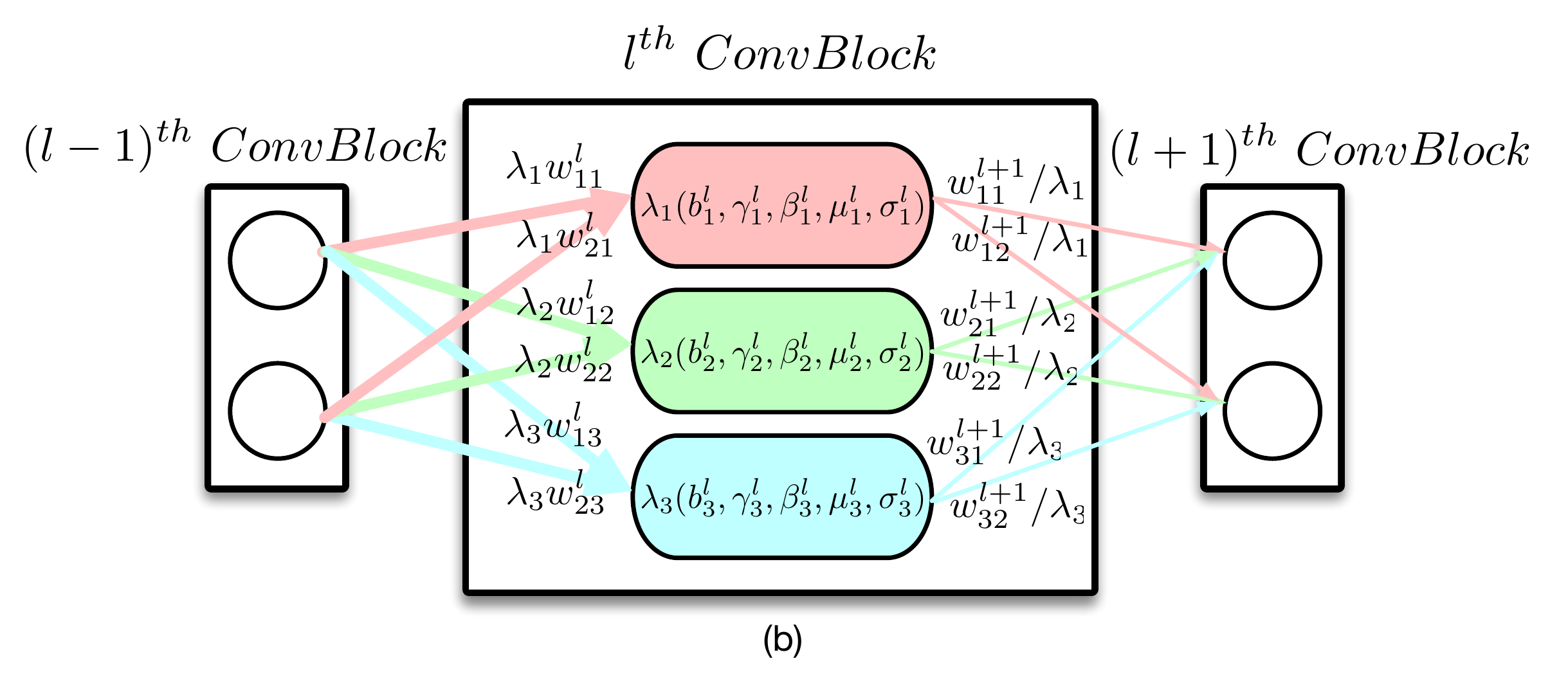}
\caption{An example of NeuronScale with $\lambda _i>0$.}
\label{fig:neuronscale}
\end{center}
\vspace{-0.3in}
\end{figure}
As the non-linear ReLU activation function is non-negative homogeneity, we can multiply the incoming weights and cancel out the effect by dividing the related outgoing weights of a random selected neurons with a positive scalar $\lambda$ simultaneously for achieving an equivalent model \cite{neyshabur2015pathsgd}. As Fig.\ref{fig:neuronscale} shows, we denote the NeuronScale operation on the neuron $n^l_i$ of model $f$ with $\lambda_i > 0$ as $\mathcal{NS}(n^{l}_i, \lambda_i)$ to obtain a new model $f'$. Formally, applying $\mathcal{NS}(n^{l}_i, \lambda_i)$ first multiplies the incoming weights with $\lambda_i$ to change the activation map of the neuron $n^l_i$ , i.e., $h^l_i{}' = h^l_i(x; \lambda_i W^l_{i,in}) = \lambda_i h^l_i$, and then divide the outgoing weights which connect the neuron $n^l_i$ with other neurons, i.e., $w^{l+1}_{ij}{}' = w^{l+1}_{ij} /\lambda_i$, to provide the same output for each neuron $n^{l+1}_j$ in the next layer.

\begin{prop}
\label{prop: neuron_scaling}
Consider a deep neural network $f$ and a new deep neural network $f'$ obtained from f by a series of NeuronScale operations $\mathcal {NS}(n_i, \lambda _i)$ to a collection of hidden neurons picked at random. Then, for every input $x$, we have $f(x) = f' (x)$.
\end{prop}

\noindent$\bullet$\textbf{ SignFlip.} 
Finally, we introduce another equivalent operation innate to the BN layers. As shown in \eqref{eq:ConvBlock_i}, we can rewrite the intermediate feature of the node $n^l_i$ in $l^{th}$ layer before passing forward to the activation function $\sigma_{ReLU}$ as follows:
\begin{equation}\label{eq:bn_combine}
\gamma^l_i \frac{w^l_{\cdot i}\odot x + b^l_i - \mu^l_i}{\sigma^l_i} + \beta^l_i = \frac{\gamma ^l_i w^l_{\cdot i}} {\sigma^l_i} \odot x + \frac{\gamma ^l_i(b^l_i - \mu ^l_i)}{\sigma^l_i} + \beta^l_i.
\end{equation}

Then, reversing the positive or negative signs of $\gamma ^l_i$, $w^l_{\cdot i}$, $b^l_i$ and $\mu ^l_i$ simultaneously can directly result in an equivalence of the target model, as the activation function would receive the same feature map as before. This interesting property allows us to arbitrarily change the signs of the kernel weights and hidden feature maps of the convolutional neuron $n^l_i$ right before BN layer, while flipping the signs of pertinent scale and mean values in the following BN layer to produce the same output features, which can be denoted as SignFlip operation $\mathcal {SF}(n^l_i)$.

\begin{prop}
\label{prop: sign_flipping}
Consider a deep neural network $f$ and a new deep neural network $f'$ obtained from f by a series of SignFlip operations $\mathcal {SF}(n^l_i)$ to a collection of hidden neurons picked at random. Then, for every input $x$, we have $f(x) = f' (x)$.
\end{prop}


\section{Attack Framework}
\label{sec:attack_framework}
Below, we present our removal attack framework against the white-box watermarking schemes, which invalidates the extraction procedure of the owner-specific information embedded in the target model with no performance overhead. Based on the set of neuron invariant transforms (\S\ref{sec:dnn_invariance}), our attack in general consists of applying a chains of random invariant transforms to entirely messes the original local features in the specific set of weights used in white-box watermarks while leaving the normal functionality intact. As almost all the known white-box watermarking schemes heavily rely on the local features of the watermark-related parameters/activations (\S\ref{sec:existing_wmk}), the perturbation caused by the invariant transformation always propagates through the mask matrix and the transformation matrix during the extraction phase and drastically cracks the final extracted watermark message.

To remove the white-box watermarks strongly relying on the intrinsic locality of DNNs, we combine the three invariant neuron transforms, i.e., LayerShuffle $\mathcal{LS}$, NeuronScale $\mathcal{NS}$ and SignFlip $\mathcal{SF}$ operations, into an integrated attack framework for maliciously perturbing the local features of the specific set of watermark-related parameters. To highlight the potential effectiveness and wide applicability to various white-box watermarking algorithms with owner-specific hyper-parameters (e.g. which layer is used to embed the model watermark), our attack framework demands no prior knowledge on either the details of white-box watermarking algorithms or the training data distribution, because these invariant neuron transforms can be applied blindly only with the full access to the victim models. Moreover, as each one of the three invariant transforms produces a model equivalence, it is plausible for the attacker to apply the transforms iteratively and compositely.


Specifically, we consider an arbitrary \textit{ConvBlock} (or \textit{LinearBlock}) layer denoted as $B = \{n_i\}_{i = 1}^N$ in the victim DNN. To remove the possible identification information encoded in this layer without any prior knowledge about the owner-specific white-box watermark algorithms, we perform three model equivalent operations for all of neurons in the layer $B$. As a result, the incoming weights $W_{i, in}$ and the outgoing weights $W_{i,out}$ of each neuron in the layer $B$ will (1) scale up/down by $\mathcal{NS}$ function with a positive real number, (2) move to other position within the current layer by $\mathcal{LS}$ function with the permutation $p$, and (3) flip the signs by $\mathcal{SF}$ functions according to the sign factor. To accelerate our attack algorithm, we take advantage of the processor-efficient parallel computation by extending the NeuronScale and the SignFlip neuron transforms to layer-level operations, which perform limited matrix multiplications on all the neurons of the specific layer by the vector-form arguments.


\begin{algorithm}
 \caption{Our watermark removal attack.}
 \label{alg:attack frame}
 \begin{algorithmic}[1]
 \renewcommand{\algorithmicrequire}{\textbf{Input:}}
 \renewcommand{\algorithmicensure}{\textbf{Output:}}
 \REQUIRE The victim model $V$.
 \ENSURE The modified model ${V'}$.
 \FOR {each layer $B$ in ${V}$}
 \STATE $N \gets$ the number of neurons in $B$
 \STATE Randomly generate permutation $p$ from 1 to $N$.
 \STATE $\mathcal{LS}(B, p)$
 \STATE Randomly generate $\lambda = \{\lambda_i | \lambda _i \in \mathbb{R}^+ \}_{i=1}^N$.
 \STATE $\mathcal{NS}(B, \lambda)$
\STATE Randomly generate $s = \{s_i | s_i \in \{-1, 1\} \}_{i=1}^N$.
\STATE $\mathcal{SF}(B, s)$
\ENDFOR \\
\RETURN $V'$
\end{algorithmic} 
\end{algorithm}

\begin{table*}[t]
  \centering
  \caption{
  Effectiveness of our attack on $9$ mainstream white-box model watermarks embedded in industry-level DNNs. Results in the \textbf{Model Utility} column report the performance of the protected model before/after our attack. In \textbf{Passport-based} rows, the values in the brackets represent the model performance with private passport or passport-aware branch.
  }
  \scalebox{0.8}{
    \begin{tabular}{lllllllll}
    \toprule
 \multirow{2}[4]{*}{\textbf{Category}}  & \multirow{2}[4]{*}{\textbf{Watermark Scheme}}  & \multirow{2}[4]{*}{\textbf{Model-Dataset}} & \multirow{2}[4]{*}{\textbf{Model Utility}} & \multicolumn{5}{c}{\textbf{Bit Error Rate (BER) of Extracted Watermarks}} \\
\cmidrule{5-9}     &  &       &       & \textit{w/o. Attack} & \textit{NeuronScale} & \textit{LayerShuffle} & \textit{SignFlip} & \textit{Unified Attack} \\
    \midrule
    \multirow{5}[10]{*}{Weight-based} & Uchida et al. \cite{uchida2017embedding} & WRN-Cifar10 & 91.55\%/91.55\%& 0   & {46.88\%} & {54.38\%} & {66.25\%} & {50.63\%} \\
\cmidrule{2-9}          & RIGA \cite{wang2021riga} & Inceptionv3-CelebA  & 95.90\%/95.90\% & 0   & {44.18\%} & {33.66\%} & {52.03\%} & {49.78\%} \\
\cmidrule{2-9}          & IPR-GAN \cite{ong2021iprgan} & DCGAN-CUB200 & 54.33/54.33 (FID) & 0   & 0 & {46.88\%} & {50\%} & {52.46\%} \\
\cmidrule{2-9}          & Greedy Residuals\cite{liu2021greedyresiduals} & ResNet18-Caltech256 & 55.05\%/55.05\% & 0   & {2.73\%} & {47.66\%} & {51.95\%} & {52.34\%} \\
\cmidrule{2-9}          & Lottery Verification\cite{chen2021lottery} & ResNet18-Cifar100 & 66.40\%/66.40\% & 0   & 0 & {48.39\%} & 0 & {48.39\%} \\
    \midrule
    \multirow{2}[4]{*}{Activation-based} & DeepSigns \cite{darvish2019deepsigns} & WRN-Cifar10 & 89.94\%/89.94\% &  0     &   {47.50}\%   &    {51.88}\%   &   {59.38}\%   & {50.63}\%  \\
\cmidrule{2-9}          & IPR-IC \cite{lim2022ipcaption} & ResNet50+LSTM-COCO & 72.06/72.06 (BLEU-1) &    0   &   0    &   {48.13}\%    &   {50}\%    & {51.25}\%\\
    \midrule
    \multirow{2}[4]{*}{Passport-based} & DeepIPR \cite{fan2021deepip} & ResNet18-Cifar100 & 67.94\%/67.94\% & 0 (67.89\%) & {48.28\% (1.13\%)} & {50.43\% (1.77\%)}  & {50\% (4.91\%)} & {52.15\% (0.30\%)} \\
\cmidrule{2-9}          & Passport-Aware \cite{zhang2020passportaware} & ResNet18-Cifar100 & 74.78\%/74.78\% & 0 (72.74\%) & {22.05\% (1\%)} & {50.54\% (1\%)} & {44.77\% (1\%)} & {51.73\% (1\%)} \\
    \bottomrule
    \end{tabular}}%
  \label{tab:eval}%
  \vspace{-0.1in}
\end{table*}%


Algorithm \ref{alg:attack frame} depicts the complete procedure of our removal attack framework. With the victim model $\mathcal{V}$ as the input, our attack algorithm outputs a modified model $\mathcal{V}'$ which is functionally equivalent to $V$ but produces a fully random message when judged by the watermarking verification algorithm. For each \textit{ConvBlock} (or \textit{LinearBlock}) $B$ in the victim model, our attack first gets the number of neurons of this layer (Line 2), denoted as $N$. Then we randomly generate three arguments in vector form of length $N$ to perform each equivalent operations, including the random permutation $p$ of integers from 1 to $N$ (Line 3), the scale factors $\lambda = \{\lambda_i | \lambda _i \in \mathbb{R}^+ \}_{i=1}^N$ (Line 5), and the sign factors $s = \{s_i | s_i \in \{-1, 1\} \}_{i=1}^N$ (Line 7). Finally, this attack algorithm will produce a model equivalence $\mathcal{V'}$.

\noindent\textbf{Attack Cost Analysis.} Based on the model invariant property to obtain an equivalence of the watermarked model, our proposed removal attack provably leaves the original model utility intact, involves no prior knowledge on the white-box watermarking algorithms or on the distribution of the original training data. Furthermore, our attack procedure requires much less computational resource as the watermark removal procedure only consists of several matrix multiplications compared with existing watermark removal attacks, which usually involves more time-consuming operations, e.g., invoking the extra training epochs to conduct fine-tuning or overwriting.

\section {Case Study}
\label{sec:case_study}
 To validate the efficiency and effectiveness of our removal attack, we make a comprehensive case study for evaluating the prevalent vulnerability of $9$ existing white-box watermark schemes from three mainstream categories, namely, weight-based, activation-based and passport-based watermarking schemes (\S\ref{sec:existing_wmk}), which are listed in Table \ref{tab:eval}.

\noindent\textbf{Evaluation Setups.} For each watermarking scheme, we strictly follow the same experiment settings from the official implementations to reproduce a watermarked model for fair evaluation. This includes but not limited to the model architecture, dataset and watermark-related hyper-parameters, on which they claim the robustness to existing removal attacks. Also, we employ the same signature $s=$``\textit{this is my signature}'' in these known watermark schemes. These methods protect the IP of diverse models, including ResNet, Inception for image classification, DCGAN for image generation and LSTM for image captioning task, which hence support the broad applicability of our proposed attack. For attack effectiveness, we use \textit{Bit Error Rate} (BER), i.e.,  the proportion of modified bits in the extracted watermark compared to the pre-defined signature,  to measure how much the watermark is tampered by our removal attack. For utility loss, we report the performance of watermarked model before/after our removal attack, i.e., FID for image generation and BLEU-1 for image captioning task and classification accuracy for other tasks \cite{ong2021iprgan,lim2022ipcaption}. We view our attack succeeds if it raises the BER of the extracted watermark to random bit strings (i.e., within a small margin of $50\%$), which ruins the ownership verification \cite{fridrich1998imagewmdetection}. We provide more detailed implementation of our attack in Appendix \ref{sec:app:config}.

 \noindent\textbf{Summary of Results.} Existing watermarking schemes once claim high robustness to some or all the previously known attacks. However, due to their heavy dependence on the element-level representation of the model parameters,  most existing watermarking schemes can no longer remain resilient against our proposed attack, yielding an almost-random watermark message after our removal attack. Table \ref{tab:eval} summarizes the results of our evaluation. In the following, we provide a case-by-case analysis on the vulnerability of existing watermark schemes and how our attack is able to crack them by manipulating the local neuron features. More evaluation results of our attack can be found in Appendix \ref{sec:app:eval}.

\subsection{Uchida et al. \cite{uchida2017embedding} (Weight-based)}
\noindent\textbf{Protection Mechanism.}
Uchida et al.\cite{uchida2017embedding} introduce the first white-box watermarking approach, which has already been used by a series of works \cite{chen2019deepmarks, chen2019deepattest}. 
During ownership verification, the authors first average the watermark-related weights $w \in \mathbb{R}^{N_{l-1}\times N_l \times k_1 \times k_2}$ through first dimension and flatten to $\hat w \in \mathbb{R}^{(N_l \cdot k_1 \cdot k_2)}$. Then, a pre-defined linear matrix $X$ and a threshold function $T_h$ at 0 further transform $\hat w$ to a binary string $s'$, i.e.,
$
s' = T_h(X\cdot \hat w),
$
which is matched with the owner-specific message $s$ in terms of BER for validation.

\noindent\textbf{Discussion.}
Although this scheme is previously shown to be vulnerable to known attacks by, e.g., overwriting the existing watermark by embedding his/her own watermarks into the target model, known attacks however demand for the specific prior knowledge of the watermark algorithm as well as the distribution of original training dataset, which are usually impractical. In this paper, we aim to disclose the underlying insecurity of this algorithm by producing an equivalence of the target model. Consider the extraction procedure of the $i^{th}$ byte $s_i'$ from the victim model, i.e. $s_i' = T_h(X_i\cdot \hat w) \in \{0,1\}$, where the linear transformation vector $X_i$ is randomly sampled from the normal distribution during embedding process. Apparently, $s_i'$ will be altered once the signs of $w$ are flipped simultaneously. Moreover, both of the local order and magnitude of $w$ have non-negligible influence on the sign of $X_i\cdot \hat w$ as $X_{ij}$ would multiply with another factor $\hat w_{p(j)}$ or scaled value $\lambda \cdot \hat w_j$.

\begin{figure}[t]
\begin{center}
\includegraphics[width=0.45\textwidth]{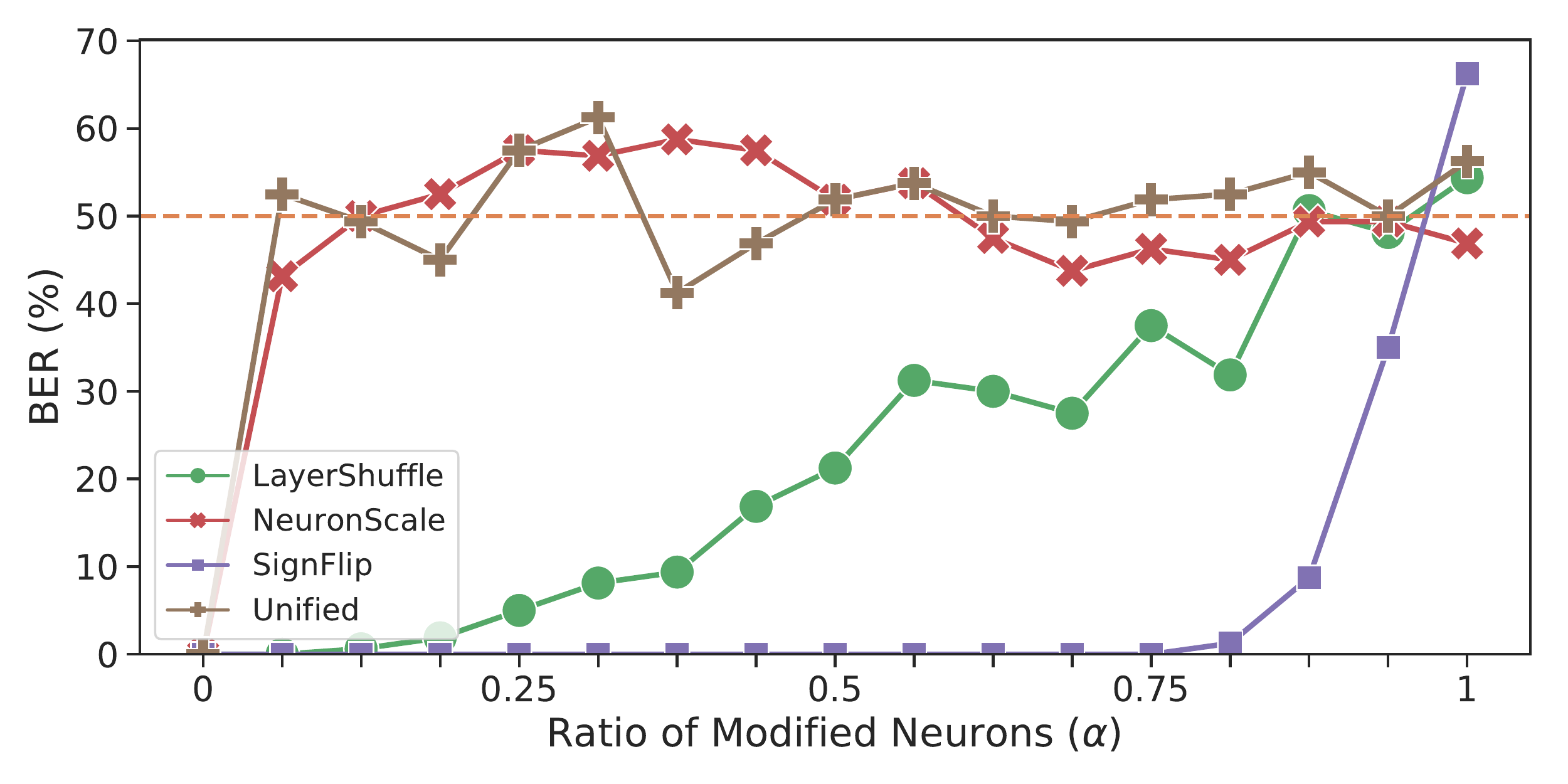}
\caption{BER of WRN watermarked by \cite{uchida2017embedding} after an $\alpha$ ratio of neurons are modified by our attack. The dashed horizontal lines reports the BER of an irrelevant model.}
\label{fig:ber_p_uchida}
\end{center}
\vspace{-0.3in}
\end{figure}

\noindent\textbf{Evaluation Results.}
We follow their evaluation settings to watermark a Wide Residual Network (WRN) trained on CIFAR10 dataset, which achieves 91.55\% accuracy and $0$ BER \cite{code-uchida}. With our attack framework, we successfully remove the watermark from the victim model without causing any degradation to the original utility. Moreover, Fig.\ref{fig:ber_p_uchida} shows applying NeuronScale or our unified attack on less than $10\%$ neurons in the victim model can completely erase the secret watermark, while LayerShuffle and SignFlip become effective removal attacks after increasing the number of modified neurons, indicating the underlying vulnerability of this algorithm.

\subsection{RIGA \cite{wang2021riga} (Weight-based)}
\noindent\textbf{Protection Mechanism.}
Based on adversarial training and more sophisticated transformation function, Wang et al. \cite{wang2021riga} improve the covertness and robustness of previous white-box watermarking techniques against watermark detection and removal attacks. They train a watermark detector to serve as a discriminator to encourage the distribution of watermark-related weights to be similar to that of unwatermarked models. Meanwhile, they replace the watermark extractor, which is originally implemented with a predefined linear transformation \cite{uchida2017embedding}, with a learnable fully-connected neural network (FCN), for boosting the encoding capacity of watermarking messages.
Similar to Uchida et al.\cite{uchida2017embedding}, the watermark-related weights are first selected from the target model and then projected to a binary string $s'$ via the FCN-based extractor (i.e., the transformation function $A$).


\noindent\textbf{Discussion.}
For strengthening the watermark covertness, RIGA sorts the watermark-related weights before training the discriminator, in consideration of the permutation invariance of neural networks. However, the invariant properties of DNNs are not properly tackled in their implementation of either the watermark embedding or extraction processes. Simply replacing the linear transformation matrix in Uchida et al. \cite{uchida2017embedding} to a learnable extractor can not completely eliminate the removal threats from our attack based on invariant transformations. As a result, RIGA algorithm is confront with the similar but critical vulnerability to \cite{uchida2017embedding} as their watermark extraction procedures only differ into the type of extractor.


\noindent\textbf{Evaluation Results.}
We run the code of RIGA they publicly released to reproduce a watermarked model of Inception-V3 trained on CelebA \cite{code-riga}. We employ the default setups that the watermark is embedded into the third convolutional layer of the target model and the extractor is a multiple layer perceptron with one hidden layer. We perform our removal attack to obtain an equivalent model which presents the same original utility with classification accuracy of $95.90\%$ while the BER during watermark verification is increased to $49.47\%$. 

\subsection{IPR-GAN \cite{ong2021iprgan} (Weight-based)}
\noindent\textbf{Protection Mechanism.}
Ong et al. present the first model watermark framework (black-box and white-box) to help intellectual property right (IPR) protection on generative adversarial networks (GANs), a family of DNN models of wide applications in real world (e.g., generating realistic images/videos). First invoking black-box verification to collect some evidence from the suspect model via remote queries, Ong et al. propose to use the white-box verification for further extracting the watermark from the specific weights of suspicious model which are accessed through the law enforcement.

Different from Uchida et al.'s \cite{uchida2017embedding}, Ong et al. propose to embed the identification information into the scale parameters $\gamma$ of the normalization layers, rather than the convolutional weights. Correspondingly, the transformation function used in watermark verification stage consists of only a hard threshold function $T_h$, which actually extracts the signs of $\gamma$ in selected normalization layers as a binary string, i.e., 
$s' = T_h(\gamma).$

\noindent\textbf{Discussion.}
We focus on the white-box part of the watermark method. Previous works have shown that the scale parameters $\gamma$ of normalization layers are more stable than the convolution weights against existing removal attacks and ambiguity attacks, as small perturbation to these watermark-related parameters would cause significant drops to the original model utility. However, as we shown in Section \ref{sec:dnn_invariance}, local features of the scale parameters can be perturbed, while leaving the model utility intact via various invariant transformations. As transformation function $T_h$ in this watermarking scheme will completely maintain the changes of the signs and order of these scale parameters, our attack successfully increases the BER of extracted binary string to $52.34\%$.

\noindent\textbf{Evaluation Results.}
We follow their evaluation setups to watermark DCGAN trained on the CUB200 dataset, which achieves $54.33$ FID and has $0$ BER \cite{code-iprgan}.
With our proposed removal attacks, the signature extracted from the scale parameters only has a $47.53\%$ match to the pre-defined binary signature while the original model utility to generate a specific image of CUB200 dataset from the Gaussian noise is perfectly preserved with $54.33$ FID. As the watermark extraction procedure including mask matrix and sign function ignores the magnitude of these scale parameters, NeuronScale has no effect on removing the secret watermark, while other two equivalences both completely confuse the ownership verification independently, as Table \ref{tab:eval} shows.

\subsection{Greedy Residuals \cite{liu2021greedyresiduals} (Weight-based)}
\noindent\textbf{Protection Mechanism.}
This method utilizes the residuals of a subset of important parameters to protect the model intellectual property. To embed the owner-specific watermark, Liu et al. \cite{liu2021greedyresiduals} greedily select fewer and more important model weights to build the residuals (i.e., \textit{Greedy Residuals}), which contributes to the decrements of the impairment caused by the updated parameters and improvements of the robustness against various watermark attacks. 

For ownership verification, Greedy Residuals extract the identity information with only simple fixed steps without either external data source or linear projection matrices. Specifically, the transformation function $A$ first applies a one-dimensional average pooling over the flattened parameters $\hat w$ in the chosen convolutional layers, and then greedily takes the larger absolute values in each row by a ratio of $\eta$ to build the residuals. Finally, the secret binary string can be extracted from the signs of residuals by hard threshold function $T_h$ after being averaged to a real-valued vector.
Formally, the extraction procedure can be written as
$
s' = T_h(Avg(Greedy(Avg\_pool\_1D(\hat w)))).
$

\noindent\textbf{Discussion.}
Greedy Residuals utilize some important parameters for embedding, which are more stable than choosing all the convolution weights in the specific layer proved in their ablation evaluations. However, the 1D average pooling and the following greedy selection of the result vector both rely on the neuron-level representations of the original weights in the watermark-related layer. Moreover, they use Rivest-Shamir-Adleman (RSA) algorithm to convert the identity information into a 256-bit sequence as the model watermark, which largely increases the length of an embeddable watermark. However, the improvement of capacity sacrifices the robustness: once the extracted watermark with few unmatched bits, they can no longer match the predefined identity information provided by the legitimate owner's private key, producing insufficient evidence for claiming the model ownership.

\noindent\textbf{Evaluation Results.}
We run the source codes of Greedy Residuals publicly released by the authors \cite{code-greedy} to reproduce a watermarked ResNet18 training on Caltech256 dataset with $55.05\%$ accuracy and $0$ BER. We first encrypt our identity information by RSA algorithm into a binary string, then we embed it on the parameters of the first convolution layer with $\eta = 0.5$. We prove that our removal attack can utterly destroy the model watermark embedded into the residual of fewer parameters, leading to an increase in the BER to $52.34\%$ on average whereas the model utility remains unchanged.

\subsection{Lottery Verification \cite{chen2021lottery} (Weight-based)}
\noindent\textbf{Protection Mechanism.}
The Lottery Ticket Hypothesis (LTH) explores a new scheme for compressing the full model to reduce the training and inference costs. As the topological information of a found sparse sub-network (i.e., the winning ticket) is a valuable asset to the owners, Chen et al. propose a watermark framework to protect the IP of these sub-networks \cite{chen2021lottery}. Specifically, they take the structural property of the original model into account for ownership verification via embedding the watermark into the weight mask in several layers with highest similarity to enforce the sparsity masks to follow certain 0-1 pattern. The proposed lottery verification uses QR code to increase the capacity of the watermark method. For watermark verification, this algorithm first selects a set of watermark-related weight masks $m$ by the mask matrix $M$ and then averages the chosen masks to a 2-dimensional matrix, which is further transformed to a QR code via $T_h$, i.e., $s' = T_h(Avg(m))$. Finally, the confidential information can be validated by scanning the extracted QR code. 

\noindent\textbf{Discussion.}
While most existing watermark techniques explore the specific model weights or prediction to embed the secret watermark, the lottery verification leverages the sparse topological information (i.e., the weight masks) to protect the winning ticket by embedding a QR code which can be further decoded into the secret watermark. However, the weight masks are closely related to the positions of the winning ticket parameters in the full model, which can be easily modified by the permutation to the relative weights. Specifically, we obtain the LayerShuffle equivalence of the trained winning ticket by generating a shuffled weight mask, which cause strong perturbation on the embedded watermark while has no change on the model performance.

\begin{figure}[t]
\begin{center}
\includegraphics[width=0.5\textwidth]{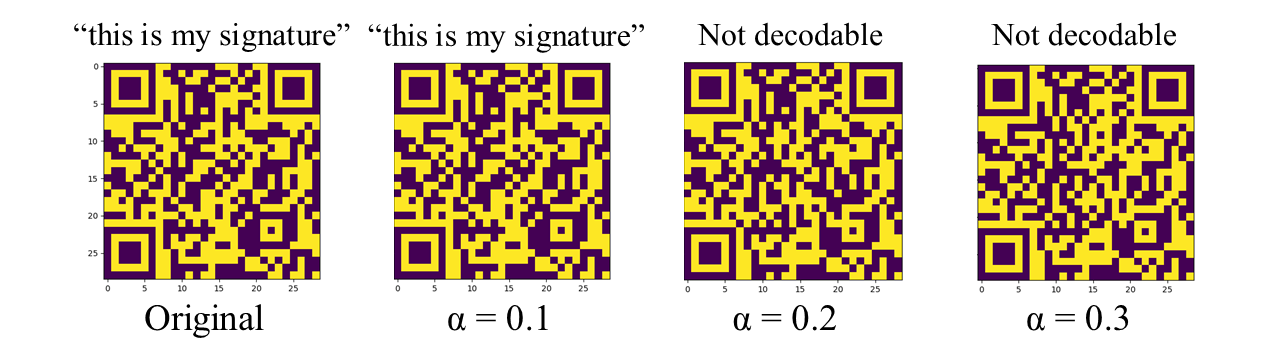}
\caption{QR code extracted from ResNet-20 watermarked by \cite{chen2021lottery} after an $\alpha$ ratio of neurons are modified by our unified attack.}
\label{fig:qrcode}
\end{center}
\vspace{-0.3in}
\end{figure}

\noindent\textbf{Evaluation Results.}
We follow the evaluation settings in the original paper to watermark a ResNet20 training on CIFAR-100 dataset, which achieves $66.41\%$ accuracy and $0$ BER \cite{code-lottery}. Although the QR code has the ability to correct some errors which improves the robustness against existing attacks, our removal method is quite effective to perturb the pattern of weight masks, invalidating the identity information decoding procedure from the extracted QR code via modifying only a few (e.g. $20\%$)  neurons in the victim model as Fig.\ref{fig:qrcode} shows.

\subsection{DeepSigns \cite{darvish2019deepsigns} (Activation-based)}
\noindent\textbf{Protection Mechanism.}
As a representative of activation-based white-box watermarking schemes, DeepSigns proposes to embed the model watermark into the probability density function (PDF) of the intermediate activation maps obtained in different layers on the white-box scenario. Specifically, DeepSigns adopts a Gaussian Mixture Model (GMM) as the prior probability to characterize the hidden representations, and considers the mean values of the PDF at specific layers to share the same role as the watermark-related weights in Uchida et al. \cite{uchida2017embedding}. Similar to the verification procedure of \cite{uchida2017embedding}, 
a transformation matrix $A$, randomly sampled in embedding procedure, projects the mean values of chosen intermediate features to a real-valued vector. With the final hard threshold function, the resulted binary string $s'$ is matched to the owner-specific watermark for claiming the model ownership. 


\noindent\textbf{Discussion.}
The notable difference between DeepSigns and \cite{uchida2017embedding} is where to embed the model watermark. 
However, as the hidden features utilized by DeepSigns are generated by the weights in the preceding layer, e.g., $a_i = W_i\cdot x+b_i$, the neuron level representations of these weights are closely related to the output feature maps. 
For example, the order of output features are shuffled concurrently after LayerShuffle operations on the watermarked layer by permutation $p$, i.e., $a_i' = W_{p(i)}\cdot x+b_{p(i)} = a_{p(i)}$.
As a result, DeepSigns is almost as vulnerable as Uchida et al's under our attack.

\noindent\textbf{Evaluation Results.}
We run the source code of DeepSigns from \cite{code-deepsign} to watermark a wide residual network trained on CIFAR10. This watermarked WRN achieves $89.94\%$ accuracy and $0$ BER. The ownership verification of the target model is completely confused by our removal attacks, as the BER is increased to $50.63\%$ with the full original model functionality.

\subsection{IPR-IC \cite{lim2022ipcaption} (Activation-based)}
\noindent\textbf{Protection Mechanism.}
As previous model watermarking schemes, which used to deploy in the classification models, are insufficient to IP protection for image captioning models and cause inevitable degradation to the image captioning performance, Lim et al. \cite{lim2022ipcaption} embed a unique signature into Recurrent Neural Network (RNN) through hidden features. In the ownership verification stage, the mask matrix $M$ first selects the hidden memory state $h$ of given watermarking image in protected RNN model. Then, the hard threshold function transforms the chosen $h$ to a binary string $s'$, which can be formally written as
$
s' = T_h(h).$

\noindent\textbf{Discussion.}
Similar to DeepSigns \cite{darvish2019deepsigns} and IPR protection on GANs \cite{ong2021iprgan}, the neuron-level features including signs and orders are closely related to the hidden memory state. Although the protected image captioning model contains an RNN architecture, the local features of neurons also suffer from invariant neuron transforms including SignFlip and LayerShuffle. For example, flipping the signs of weights related to the chosen $h$ in RNN cells and the weights in the dense layer, which decodes the outputs of RNN into words, produces an equivalence of the original model with the same output as \textit{tanh} in RNN is an odd function. More details can be found in Appendix \ref{sec:app:coverage}.

\noindent\textbf{Evaluation Results.}
We run the official implementation \cite{code-captioning} to reproduce a watermarked Resnet50-LSTM trained on COCO, which achieves $72.06$ BLEU-1 and has $0$ BER. With our proposed removal attack, the signature extracted from the hidden memory state $h$ has $51.25\%$ BER compared to the owner-specific binary message, while no loss brings to the image captioning performance. As the signature extracted from the signs of $h$, NeuronScale is unable to increase the BER in verification stage, while the other two equivalences both show the strong effectiveness to the watermark removal (Table \ref{tab:eval}).

\subsection{DeepIPR \cite{fan2021deepip} (Passport-based)}\label{section:DeepIPR}
\noindent\textbf{Protection Mechanism.} 
DeepIPR is one of the earliest passport-based DNN ownership verification schemes \cite{fan2021deepip}. By inserting an owner-specific passport layer during the watermark embedding procedure, DeepIPR is designed to claim the ownership not only based on the extracted signature from the specific model parameters but also on the model inference performance with the private passport layer. Consequently, this scheme shows high robustness to previous removal attacks and especially to the ambiguity attacks, which mainly forges counterfeit watermarks to cast doubts on the ownership verification.

In our evaluation, we focus on the following passport verification scheme in \cite{fan2021deepip}. This scheme generates two types of passport layers simultaneously by performing a multi-task learning, i.e., public passports for distribution and private passports for verification, both of which are actually based on normalization layers. Generally, DeepIPR leverages pre-defined digital passports $P=\{P_{\gamma}, P_{\beta}\}$ to obtain the scale and the bias parameters of the private passport, which are written as:
$
\gamma = Avg(W_c \odot P_{\gamma}), \beta =  Avg(W_c \odot P_{\beta}),
$
where $W_c$ is the filters of the precedent convolution layer, and $\odot$ denotes the convolution operation. DeepIPR adopts a similar watermark extraction process as \cite{uchida2017embedding}, where the transformation function $A$ converts the signs of the private $\gamma$ into a binary string to match the target signature.

\noindent\textbf{Discussion.}
In their original evaluation on the DeepIPR's robustness, the private passport encoded with correct signature achieves a similar classification accuracy to the public passport against the existing removal attacks. By their design, forging new passport layers is particularly tough due to the close link between an accurate watermark and the non-degenerate model performance. However, as the private $\gamma$ and $\beta$ are obtained from the preceding convolution weights, DeepIPR actually embeds the secret signature in the hidden output of the convolutional layer with the weights $W_c$ given the input $P_{\gamma}$ or $P_{\beta}$. Similar to the activation-based watermarking scheme, the neuron-level representations of the related parameters are crucial to the BER of extracted signature. Moreover, the equivalent transformations can invalidate the private passport, as the private $\beta$ would not align to the transformed model any longer. We provide more analysis and proof of this performance degradation in Appendix \ref{sec:app:passport}.

\begin{figure}[t]
\begin{center}
\includegraphics[width=0.45\textwidth]{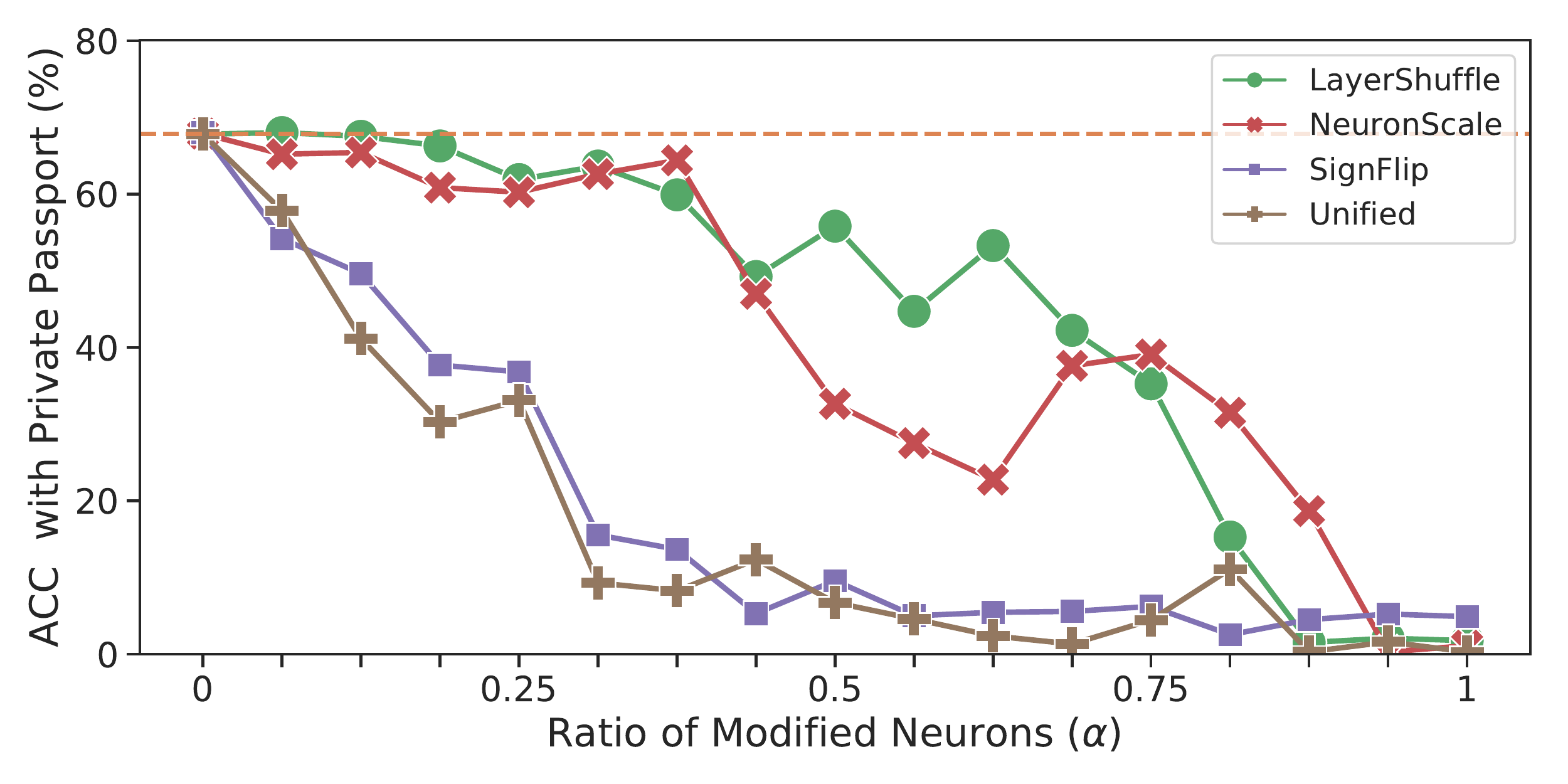}
\caption{Performance of ResNet-18 (watermarked by DeepIPR) with the private passport when an $\alpha$ ratio of neurons are modified by our attacks, where the dashed horizontal line reports the original performance.}
\label{fig:passport_deepipr}
\end{center}
\vspace{-0.3in}
\end{figure}

\noindent\textbf{Evaluation Results.}
We evaluate our attack on the watermarked ResNet18 trained on the CIFAR-100 dataset with DeepIPR \cite{code-deepipr} which achieves $67.94\%$ accuracy with public passport and  $67.89\%$ accuracy with private passport, respectively. As DeepIPR replaces the BN layers in ResNet18 to GN layers, we adapt our attack algorithms for NeuronScale and SignFlip which are presented in Appendix \ref{sec:app:coverage}.
As Fig.\ref{fig:passport_deepipr} shows, our proposed removal attack successfully decreases the performance of the model with private passports to $0.30\%$ and increases the BER to $52.15\%$ after modifying all the weights in the protected model (i.e. $\alpha = 1$), while causing no change in the accuracy of the original model with the public passports. 

\subsection{Passport-aware Normalization \cite{zhang2020passportaware} (Passport-based)}
\noindent\textbf{Protection Mechanism.} 
Zhang et al. \cite{zhang2020passportaware} propose another passport-based watermark method without modifying the target network structure, which would otherwise incur notable performance drops. They adopt a simple but effective strategy by training the passport-free and passport-aware branches in an alternating order and maintaining the statistic values independently for the passport-aware branch at the inference stage. Similar to DeepIPR, the authors design the learnable $\gamma, \beta$ to be relevant to the original model for stronger ownership claim. During the extraction of model watermarks, the transformation function $A$ first projects the $\gamma$ by an additional FCN model to an equal-length vector and then utilizes the signs of the vector to match the target signature.

\noindent\textbf{Discussion.}
While this method improves DeepIPR in terms of model performance by preserving the network structure, we discover it still relies on the same set of local neuron features as in DeepIPR.  Even the improved transformation function $A$ in this method including linear transformation and sign function can hardly block the modification on neuron-level representations of the watermark-related parameters (i.e., the convolution weights before the passport-aware branch).


\begin{table}[htbp]
  \centering
   \caption{Extracted signatures and the model accuracy when an $\alpha$ ratio of neurons are modified by our attack. Values in the bracket report the model accuracy with the passport-aware branch.}
    \begin{tabular}{clcc}
    \toprule
    \textbf{$\alpha$} & \textbf{Signature s} & \textbf{BER (\%)} & \textbf{Accuracy (\%)} \\
    \midrule
    0     &$this\ is\ my\ signature$ &  0     & 74.68 (72.73) \\
    1/16  & ôhisióíqóùgîaôõe & 8.04  & 74.68 (1.00) \\
    1/8   & ÔHéSÉñ@YÓëN¦AôEòE & 16.46  & 74.68 (1.00) \\
    1/4   &XîEw.oN\&!\$õuaF÷zã~a &  30.28  & 74.68 (1.00) \\
    1/2   & ±*:ãâ\%W3Ú,dE & 45.83  & 74.68 (1.00) \\
    1     & @zs~ó6Ë6áÓ=ñUÇ & 50.96  & 74.68 (1.00) \\
    \bottomrule
    \end{tabular}%
  \label{tab:sig1}%
\end{table}%


\noindent\textbf{Evaluation Results.}
When we embed the model watermark into a ResNet18 trained on the CIFAR-100 via passport-aware normalization \cite{code-aware}, we are able to achieve $0$ BER, while preventing the original model utility from unacceptable drops.
As Table \ref{tab:sig1} shows, when we gradually modify a certain proportion of neurons with our attack, the signature extracted from the victim model becomes increasingly unreadable, especially when $\alpha \ge {1}/{8}$. Specifically, LayerShuffle and SignFlip demonstrate sufficient effectiveness to remove this white-box watermark and invalidate the passport-aware branch independently, while NeuronScale boosts the BER to $22.05\%$ and turns the model performance with passport-aware branch from $72.74\%$ to  random guess (i.e., $1\%$).



\section{Discussions}\label{sec:discussion}


\subsection{Coverage of Victim Models}
\label{sec:coverage}
First, we discuss the broad applicability of our attack based on invariant neuron transforms on other architectural options of DNN models. The rigorous analysis and proofs for the broader applicability can be found in Appendix \ref{sec:app:coverage}.


\noindent\textbf{Applicable to More Normalization Layers.}
Similar to the case of batch normalization we can obtain a SignFlip equivalence of the target model with group normalization, instance normalization or layer normalization by only reversing the signs of learnable parameters in the normalization layer and the preceding convolution/linear layer. Although the statistic values of these normalization layers are calculated on the fly during inference and the adversary is unable to modify them directly, the modifications on the trainable parameters of the preceding layers already result in sign flips in these calculated statistic values to produce a equivalent model, which is also corroborated by our evaluation results in Section \ref{section:DeepIPR}. 

\noindent\textbf{Applicable to More Activation Functions.} Similar to the case of ReLU, our attacks are applicable to the DNNs with other activation functions such as LeakReLU, PReLU, RReLU and \textit{tanh} \cite{he2015rectifiers,xu2015linearunit}, only if the activation function $\sigma$ satisfies $\sigma (ax) = a\sigma(x)$ with $a\in \mathbb{R}$ in testing. We admit some other activation functions, e.g., \textit{sigmoid}, would disable the equivalence under NeuronScale. Nevertheless, it has limited effect on the effectiveness of our unified attack framework when attacking DNN architectures with these special activation functions, as the other two transforms (i.e., SignFlip and LayerShuffle) are able to remove the potential watermark regardless of the choice of activation functions.

\noindent\textbf{Applicable to Other Complex Model Architectures.}
Besides, our removal attack can be easily applied to the watermarked DNNs with special connections between the neurons of different layers, e.g. ResNet with shortcuts \cite{he2016resnet} and Inception-V3 with parallel convolution \cite{szegedy2016inception} operations, which have much more complicated architecture than the simple convolutional neural networks. For example, we can obtain the equivalent branches in each Inception block separately to remove the embedded watermark in Inception-V3. Only if the adversary knows well about the forwarding computation in the target DNN, which is a common knowledge in white-box watermarking settings, he/she can readily extend our attack with small adjustments, which is left for future works. 

\noindent\textbf{Revealing the Vulnerability of Other Works.}
During our extensive literature research, we surprisingly discover that not only the white-box watermarking schemes but also other works overly depend on the specific local model features which are vulnerable to our proposed attack. For example, Chen et al \cite{chen2021copy} present DEEPJUDGE, a testing platform based on non-invasive fingerprint for copyright protection of DNNs, which measures the similarity between the suspect model and victim model using multi-level testing metrics. Despite the authors claiming strong robustness against various attacks in their evaluation, we notice that $4$ of the $6$ testing metrics in DEEPJUDGE measure the distance between the two models' hidden layer/neuron outputs in the white-box scenario, ignoring the existence of equivalent victim models which can be produced by invariant neuron transforms. For example, LayerShuffle would cause the outputs of the transformed model to be no longer be aligned with the outputs of the original victim model, which may therefore invalidate the copyright statement. We strongly suggest existing defensive works involving DNNs further rethink the robustness when invariant neuron transforms are available to attackers.


\subsection{Black-box vs. White-box Watermarking Algorithms}
Existing black-box watermark schemes \cite{adi2018turning, jia2021entangled} mostly embed the identity information into the input-output patterns of the target model on a secret trigger set (similar to backdoor attacks \cite{gu2017badnets}). This inevitably degrades the original model utility. As reported in \cite{adi2018turning}, the trade-off is sometimes evident between successfully embedding a black-box watermark and correctly producing predictions on normal inputs. Moreover, recent progresses on backdoor defenses also expose a new attack surface of these black-box watermark techniques \cite{liu2019abs, liu2018finepruning, wang2019neuralcleanse}. Specifically, the adversary can either remove the secret watermark via performing specific transforming to the trigger data or throw doubt on the security of the watermarked model by successfully detecting some trojaned neurons.

As the black-box and white-box watermarking schemes do not conflict with each other, some recent works combine them to provide more robust protection to the model IP \cite{chen2021lottery,fan2021deepip,zhang2020passportaware,ong2021iprgan,darvish2019deepsigns}. During their watermark verification, these hybrid watermark algorithms first collect sufficient evidence via remote queries to the suspect models. Then, the owner further attains the full access of the model with law enforcement to detect the identity information in the model internals, which yields a strong copyright statement. As our attack framework breaks most of the existing verification procedures in white-box watermark algorithms, these hybrid watermark schemes could hardly survive under our attack.

Recently and concurrently with our work, Lukas et al. at S\&P'22 \cite{sokwatermark} also evaluate whether existing watermarking schemes are robust against known removal attacks. Our work reveals the vulnerability of the white-box watermarking schemes which claim high robustness, while Lukas et al. mainly focus on black-box ones. Although they demonstrate a preliminary feature permutation attack against DeepSigns \cite{darvish2019deepsigns}, they incorrectly claim that other white-box watermarks (e.g., \cite{uchida2017embedding}) are permutation-invariant. It is a false sense of robustness. We prove that shuffling the weights from the first two dimensions by blindly applying our proposed LayerShuffle still cracks the victim model (\S\ref{sec:case_study}). Based on a systematic study of invariant neuron transforms and their implication on the security of white-box watermarks, our work requires no prior knowledge to the watermark algorithms and training data distribution, while causing no utility loss. However, the requirements on prior information and a clear performance drop are proven to be inevitable in their evaluation. 

\subsection{Extension of Our Attack}
By converting random invariant neuron transforms to particular ones, our proposed attack is readily for \textit{watermark overwriting}, which implants the identity information of malicious users. For example, we could incorporate any binary signature into the specific scale parameters of a victim model watermarked by DeepIPR. Specifically, the adversary can force the signs of scale weights of selected layers to match the specified signature, while preventing the original model utility from any unexpected impairment by reversing the signs of other relative parameters to obtain a SignFlip equivalence of the victim model. As a result, the legitimate owner is unable to claim the ownership of the victim model inserted with the owner's private passports due to the unmatched poor accuracy and the high BER, whereas the adversary is able to extract his/her own signature from the suspect model, perfectly matching with the adversary's identity information. 

As a final remark, we highlight that the extension of our removal attack to an effective overwriting attack is a post-processing technique at almost no cost, while previous works \cite{wang2019overwrite} overwrite the existing watermark by running the watermark algorithm again to embed a new watermark. This requires unrealistic access to prior knowledge such as the original training set and details of the watermark schemes, along with additionally more computing resources.

\subsection{Potential Mitigation Strategies}
As our removal attack produces an equivalence of the victim model, a straightforward defense is to recover the original model before watermark verification by inverting the applied neuron transforms. However, our unified attack framework consists of a chain of different neuron transforms, which makes the reverse-engineering a non-trivial task. Moreover, our removal attack can combine with any post-processing techniques such as fine-tuning and pruning, which further increases the cost to recover the original model with owner-specific watermarks.  Considering the core vulnerability of most known white-box algorithms, future works on robust watermark algorithms should consider the robustness under invariant neuron transforms into account. For example, it is plausible to sort the chosen weights before calculating the watermark-related regularization loss during the watermark embedding phase. As a result, the owner can reorder the LayerShuffle equivalence of the victim model to detect their identity information. It is worth to note that a potential mitigation on the vulnerability should not sacrifice the utility of the original model and the reliability of the watermark algorithm, which indeed ensures minimal false positive in verification but to some degree intensifies the challenges of mitigating our attack. We hope future works would devise robust white-box algorithms against our removal attack.


\section{Conclusion}
By thoroughly analyzing the protection mechanisms of existing white-box model watermarks, we reveal their common yet severe vulnerability under invariant neuron transforms, which our work intensively exploits to arbitrarily tamper the internal watermark while preserving the model utility. Through extensive experiments, we successfully crack 9 state-of-the-art white-box model watermarks which claim robustness against most previous removal attacks, but are reduced to be almost random under our attack when no loss of model utility is incurred and no prior knowledge of the training data distribution or the watermark methods is required. We urgently alarm future works on white-box model watermarks to evaluate the watermark robustness under invariant neuron transforms before deploying the watermark protocol in real-world systems. Moreover, it would be very meaningful to search for more robust and resilient internal model features, which should be at least invariant under our exploited three types of invariant transforms, for model watermarking.


\bibliographystyle{IEEEtran}
\bibliography{ref}

\appendix
\subsection{Omitted Proofs for Invariant Neuron Transforms}
\label{sec:app:proof}
We first present the rigorous proofs to three invariant neuron transforms on the fully connected networks (FCNs) and simple deep convolutional neural networks (CNNs), with batch normalization layer and ReLU activation in each \textit{LinearBlock} or \textit{ConvBlock} layer. As a fully connected layer in DNNs is a simplified form of convolutional layer, we focus on showing the
invariant property in a \textit{ConvBlock}. For the $i^{th}$ neuron in the $l^{th}$ layer of the target model, as shown in Section \ref{sec:dnn_invariance}, the associated parameters can be grouped as incoming weights $W^l_{i,in} = \{\{w^l_{r,i}\}^{N_{l-1}}_{r=1}, b^l_i, \gamma^l_i, \beta^l_i, \mu^l_i, \sigma^l_i\}$ and outgoing weights $W^l_{i, out} = \{w^{l+1}_{i,j}\}^{N_{l+1}}_{j=1}$, where the output is denoted as $h^l_i(x; W^l_{i,in})$. As the outgoing weights of the $l^{th}$ layer is actually the convolutional weights in the $(l+1)^{th}$ layer, the output of the $(l+1)^{th}$ layer can be written as:
$
f^{l+1}(f^l;\{W^{l+1}_{i, in}\}^{N_l}_{i =1}) = \{max \cdot \sigma (\gamma^{l+1}_i \frac{W^{l}_{i,out}\odot f^l + b^{l+1}_i - \mu^{l+1}_i}{\sigma^{l+1}_i} + \beta^{l+1}_i)\}^{N_{l+1}}_{i=1},
$
where $f^l = \{h^l_i\}^{N_l}_{i = 1}$ is derived by Equation (\ref{eq:ConvBlock_i}).

\begin{proof}[Proof for Proposition \ref{prop: layer_shuffling}]
With a LayerShuffle operation $\mathcal{LS}(f^l, p)$ on the $l^{th}$ layer of the target model $f$ to get a new model $f'$, the permutation $p=\{p(i)\}_{i=1}^{N_l}$ reorders the incoming weights and outgoing weights to $W^l_{i,in}{}' = \{\{w^l_{r,p(i)}\}^{N_{l-1}}_{r=1}, b^l_{p(i)}, \gamma^l_{p(i)}, \beta^l_{p(i)}, \mu^l_{p(i)}, \sigma^l_{p(i)}\} = W^l_{p(i),in}$ and $W^l_{i, out}{}' = \{w^{l+1}_{{p(i)},j}\}^{N_{l+1}}_{j=1}=W^{l}_{p(i), out}$, respectively. As a result, $h^l_i{}' = h^l_i(x;W^l_{p(i),in})=h^l_{p(i)}$ and then we have $f^l{}' = \{h^l_i{}'\}^{N_l}_{i = 1}=\{h^l_{p(i)}\}^{N_l}_{i = 1}$. For the $(l+1)^{th}$ layer, $\mathcal{LS}(f^l, p)$ only shuffles the order of $W^{l+1}$ in the first dimension, which is actually the outgoing weights of the $l^{th}$ layer. Then, the intermediate result $W^{l}_{i,out}{}' \odot f^l{}' $ of the $(l+1)^{th}$ layer, which can be written as $W^{l}_{p(i), out}\odot\{h^l_{p(i)}\}^{N_l}_{i = 1}$, is exactly equal to $W^{l}_{i,out} \odot f^l$  produced by the original model, leaving the output of the $(i+1)^{th}$ layer intact, i.e., $f^{l+1}{}' = f^{l+1}$. As this LayerShuffle operation does not modify the other parameters in the following layers, we have $f(x) = f'(x)$ for every $x$.
\end{proof}

\begin{proof}[Proof for Proposition \ref{prop: neuron_scaling}]
With the NeuronScale operation $\mathcal{NS}(n^l_i, \lambda_i)$ on each neuron in the $l^{th}$ layer of the target model $f$ to get a new model $f'$, the float-point vector $\lambda=\{\lambda _i\}_{i=1}^{N_l}$ rescales the incoming weights and the outgoing weights to $W^l_{i,in}{}' = \{\{\lambda_i w^l_{r,i}\}^{N_{l-1}}_{r=1}, \lambda_i b^l_{i}, \lambda_i \gamma^l_{i}, \lambda_i \beta^l_{i}, \lambda_i \mu^l_i, \lambda_i \sigma^l_i\} = \lambda_i W^l_{i,in}$ and $W^l_{i, out}{}' = \{w^{l+1}_{{i},j}/ \lambda_i \}^{N_{l+1}}_{j=1}=  W^{l}_{i, out} / \lambda_i $, respectively. As a result, $h^l_i{}' = h^l_i(x;\lambda_i W^l_{i,in})=\lambda_i h^l_{i}$ and then we have $f^l{}' = \{h^l_i{}'\}^{N_l}_{i = 1}=\{\lambda_i h^l_{i}\}^{N_l}_{i = 1}$. For the $(l+1)^{th}$ layer, $\mathcal{NS}(n_i^l, \lambda)$ only scales down $W^{l+1}$, which is actually the outgoing weights of the $l^{th}$ layer. Then, the intermediate result $W^{l}_{i,out}{}' \odot f^l{}' $ of the $(l+1)^{th}$ layer, which can be written as $(W^{l}_{i, out} / \lambda_i )\odot\{\lambda_i h^l_{p(i)}\}^{N_l}_{i = 1}$, is exactly equal to $W^{l}_{i,out} \odot f^l$  produced by the original model, leaving the output of the $(i+1)^{th}$ layer intact, i.e., $f^{l+1}{}' = f^{l+1}$. As this NeuronScale operation does not modify the other parameters in the following layers, we have $f(x) = f'(x)$ for every $x$.
\end{proof}

\begin{proof}[Proof for Proposition \ref{prop: sign_flipping}]
With the SignFlip operation $\mathcal{SF}(n_i^l)$ on each neuron in the $l^{th}$ layer of the target model $f$ to get a new model $f'$, the signs of most incoming weights (i.e., $\gamma ^l_i$, $w^l_{\cdot i}$, $b^l_i$ and $\mu ^l_i$) are flipped, which can be written as $W^l_{i,in}{}' = \{\{-w^l_{r,i}\}^{N_{l-1}}_{r=1}, -b^l_{i}, -\gamma^l_{i},  \beta^l_{i}, -\mu^l_i, \sigma^l_i\}$. As a result, $h^l_i{}' = h^l_i(x;W^l_{i,in}{}')= h^l_{i}$ and then we have $f^l{}' = \{h^l_i{}'\}^{N_l}_{i = 1}=\{h^l_{i}\}^{N_l}_{i = 1}$, which is exactly equal to $f^l$ produced by the original model, leaving the output of the $i^{th}$ layer intact. As this SignFlip operation does not modify the other parameters in the following layers, we have $f(x) = f'(x)$ for every $x$.
\end{proof}

\subsection{Omitted Details on Attack Extension}
\label{sec:app:coverage}
In Section \ref{sec:coverage}, we discuss the broad applicability of our attack on other watermarked models with various architecture, not limited to simple convolutional neural network with ReLU activations and batch normalization layers. We provide more detailed analysis and proofs in this section.

\noindent\textbf{Applicable to More Normalization layers.}
For group normalization, the neurons in the $l^{th}$ layer (i.e., $\{n^{l}_i\}^{N_l}_{i = 1}$) are first divided into $G$ groups. Formally, the $k^{th}$ group can be denoted as  $\{n^{l}_i\}^{k* N_l / G}_{i = (k-1) * N_l / G+1}$. Different from BN, GN computes the statistic values for each group independently throughout both the training and testing phases. Similar to Eq. (\ref{eq:bn_combine}), the output of the $i^{th}$ neuron which belongs to the $k^{th}$ group can be formally written as:
\begin{equation}\label{eq:gn_combine}
\gamma^l_i \frac{w^l_{\cdot i}\odot x + b^l_i - \mu^l_k}{\sigma^l_k} + \beta^l_i = \frac{\gamma ^l_i w^l_{\cdot i}} {\sigma^l_k} \odot x + \frac{\gamma ^l_i(b^l_i - \mu ^l_k)}{\sigma^l_k} + \beta^l_i,
\end{equation}
where $\mu_k$ and $\sigma_k$ are the mean and variance of the $k^{th}$ group. 

To obtain a SignFlip equivalence, we only reverse the signs of $\gamma_i$, $w^l_{\cdot i}$ and $b^l_i$ in each neuron. Then, we have $\mu_k' = -\mu_k$ and $\sigma_k' = \sigma_k$ by definition. Finally, after our SignFlip operation, the output of each neuron with group normalization is exactly same as the original result.

To obtain a LayerShuffle equivalence, we can first shuffle the neurons within each group and then further reorder these $G$ groups within each layer. Specifically, for the $k^{th}$ group in the $l^{th}$ layer, we randomly generate permutation $p_{k,in}$ from $1$ to $N_l / G$ and permutation $p_{l,out}$ from $1$ to $G$. Then, we obtain the final permutation $p$ for the $l^{th}$ layer as $p(i) =(p_{l,out}(k)-1)N_l/G+ p_{k,in}(r)$, where the $i^{th}$ neuron belongs to the $k^{th}$ group and $i = (k-1)N_l/G+r$. With permutation $p$, we have $\mu_k' = -\mu_{p_{l,out}(k)}$ and $\sigma_k' = \sigma_{p_{l,out}(k)}$ by definition. As a result, we can transform the model with GN layers by $\mathcal{LS}(B^l, p)$ as is shown in Section \ref{sec:layershuffle} to obtain a LayerShuffle equivalence. 

To obtain a NeuronScale equivalence, we scale up/down the incoming/outgoing weights of each neuron which belongs to the $k^{th}$ group in the $l^{th}$ layer with the same positive factor $\alpha_k$. Then we leverage $\mathcal{NS}(n^l_i, \alpha_k)$ for the $i^{th}$ neuron in $k^{th}$ group to obtain a NeuronScale equivalence.

Moreover, instance normalization and layer normalization are both the simplified forms of group normalization, once we set the $G$ in group normalization to $N_l$ and $1$, respectively. As a result, our attack based on invariant neuron transforms can be further applied to the model with these three normalization layers, i.e., GN, IN and LN.

\noindent\textbf{Applicable to More Activation Functions.}
We assume the activation function in most victim models as ReLU for convenience, as NeuronScale leverages the non-negative homogeneity, i.e. $\sigma_{ReLU}(\lambda x) = \lambda \sigma_{ReLU}(x) $ for every $\lambda > 0$, which is also presented in the other kinds of rectified units, i.e., leaky rectified linear units (Leaky ReLU), parametric rectified linear units (PReLU) and randomized rectified linear units (RReLU). Formally, these activation functions can be formulated as $\sigma(x) = ax$, where $a=1$ for every $x\geq 0$ otherwise $a > 0$ remains fixed in testing. Similarly, we can set the positive factors $\lambda$ to leverage the NeuronScale operations, as the scale ratio $\lambda$ of features is preserved through these linear units, i.e., $\sigma(\lambda x) = \lambda\sigma(x) $.

Moreover, the activation function of \textit{tanh}, which is widely deployed in RNN models, has the property that $tanh(x) = tanh(-x)$ for all $x\in \mathbb{R}$. We can utilize it to obtain a SignFlip equivalence of the victim model by $\mathcal{NS}(n_i, -1)$, as the signs of the incoming and the outgoing weights of the neuron $n_i$ are reversed while the original model utility is perfectly preserved.

\noindent\textbf{Applicable to Other Complex Model Architectures.}
As most victim models in Section \ref{sec:case_study} are more complicated than \textit{ConvBlock}, we adapt our attack algorithm by carefully setting $p, \lambda, s$ for each layer.

For ResNet \cite{he2016resnet}, we consider the $l^{th}$ residual block $R^l$ which usually consists of two \textit{ConvBlock} (denoted as $B^l_1, B^l_2$) and a skip connection $S^l$ between the input $x^l$ and the output $y^l$ of this residual block. The forward propagation of $R^l$ can be formulated as: $
y^l = B^l_2(B^{l}_{1}(x^{l})) + S^{l}(x^{l}),$
where the skip connection $S^l$ is an identity function or a \textit{ConvBlock}.
As only modifying the inner weights of the target \textit{ConvBlock}, the SignFlip operation can be directly employed to ResNet with shortcuts layer-by-layer. LayerShuffle and NeuronScale on residual block are quite different due to the skip connection $S^l$. As the modified output of the first \textit{ConvBlock} $B^l_1$ can be calibrated (e.g., re-ordering or re-scaling) by the weights in the second \textit{ConvBlock} $B^l_2$, LayerShuffle and NeuronScale are applicable to $B^l_1$ with no adjustment. Once we apply LayerShuffle and NeuronScale on $B^l_2$, the output of shortcut $S^l$ should be transformed in the same way to align the features from $B^l_2$ in the forward computation of a residual block. As a result, we set the same $p$ and $s$ for the second \textit{ConvBlock} and the shortcut of each residual block when they have the same number of neurons, yielding LayerShuffle and NeuronScale equivalence.

For Inception-V3 \cite{szegedy2016inception}, the inception modules apply multiple sizes of kernel filters to extract multiple representations. We consider the $l^{th}$ inception module $I^l$ in Inception-v3 which usually consists of several branches $B^L_i$, where each branch $B^l_i$ is a sequence of \textit{ConvBlock}. Then, for a given input $x^l$, the forwarding computation of $I^L$ with $n$ branches can be written as: $
y^l = Concatenate(B^l_1(x^l), B^l_2(x^l), ..., B^l_n(x^l)).
$

Similarly, SignFlip can be directly applied to the Inception model layer-by-layer. As the output features of the branches are concatenated with each other rather than added elementwise, we only take care of the outgoing weights to get LayerShuffle and NeuronScale equivalence. For each $B^l_i$ (not last layer) in the $i^{th}$ branch, we can shuffle or scale the weights independently. For the last \textit{ConvBlock} in each branch, the outgoing weights spread into a few branches if following another inception module. Then, we can obtain the corresponding invariant models by LayerShuffle and NeuronScale.
\subsection{Omitted Analysis for the Degradation of Passport Utility}
\label{sec:app:passport}
The watermark schemes \cite{fan2021deepip,zhang2020passportaware} are designed to claim the ownership by both the low BER of extracted signature and the high model utility of private passports. As is shown in Section \ref{sec:case_study}, $\gamma$ and $\beta$ in the private passport are closely related to the convolution weights in the precedent convolution. To analyze the degradation of the model performance with private passport after our attack, we consider the $l^{th}$ \textit{ConvBlock} in victim model. Formally, the output of the private passport in $B^l$ can be written as:
\begin{equation}
\label{eq:passport}
y^l_i = \gamma_{i,pri}^l\frac{w^l_{\cdot i}\odot x^l + b_i^l - \mu_{i,pri}^l}{\sigma_{i,pri}^l} + \beta_{i,pri}^l,
\end{equation}
where $\gamma_{i,pri}^l$, $\beta_{i,pri}^l$, $\mu_{i,pri}^l $ and $\sigma_{i,pri}^l$ are the parameters in the private passport. 

\noindent\textbf{DeepIPR.}
In this watermark scheme, the normalization layers are replaced by GN and the learnable parameters in private passport are generated by the digital passports, i.e., $\gamma_{i,pri}^l = Avg(w^l_{\cdot i} \odot P^l_{\gamma}) $, $\beta_{i,pri}^l=Avg(w^l_{\cdot i} \odot P^l_{\beta})$. In the following, We proof the degradation of the model with private passports through the unexpected and irreversible modification on the $\gamma_{i,pri}$ and $\beta_{i,pri}$ after our attack.

To obtain SignFlip equivalence,  $\mathcal{SF}(n^l_i)$ only reverses the signs of $\gamma ^l_i$, $w^l_{\cdot i}$ and $b^l_i$ as $\mu ^l_i$ and $\sigma ^l_i$ in the GN are calculated on the fly, which can be proved to preserve the original model utility. As a result, the parameters in the private passport are transformed correspondingly. For example, the scale weight is generated as follows: 
\begin{equation}
\gamma_{i,pri}^l{}' = Avg(w^l_{\cdot i}{}' \odot P^l_{\gamma}) = Avg(-w^l_{\cdot i} \odot P^l_{\gamma})= -\gamma_{i,pri}^l.
\end{equation}
Similarly, we have $\beta_{i,pri}^l{}' = - \beta_{i,pri}^l$, $\mu_{i,pri}^l{}' =  -\mu_{i,pri}^l$ and $\sigma_{i,pri}^l{}' = \sigma_{i,pri}^l$ in private passport after the SignFlip operation. Then, the output of transformed private passport is altered as:
\begin{align}
y^l_i{}' &= -\gamma_{i,pri}^l \frac{-w^l_{\cdot i}\odot x^l + b_i^l -(- \mu_{i,pri}^l)}{\sigma_{i,pri}^l} - \beta_{i,pri}^l\\
&= \gamma_{i,pri}^l\frac{w^l_{\cdot i}\odot x^l + b_i^l - \mu_{i,pri}^l}{\sigma_{i,pri}^l} - \beta_{i,pri}^l \neq -y^l_i,
\end{align}
where the sign of $\beta_{i,pri}^l$ is reversed unexpectedly which can not be calibrated by transforming other model parameters. 

For LayerShuffle transformation on $l^{th}$ \textit{ConvBlock} in victim model, $\mathcal{LS}(B^l, p)$ permutes the order of incoming and outgoing weights of each neuron to preserve the original model performance, i.e. $w^l_{c, i}{}' = w^l_{c, p(i)}$ , $b^l_i = b^l_{p(i)}$, and $w^{(l+1)}_{i,j}{}' = w^{(l+1)}_{i,p(j)}$. Then, the parameter in private passport of the $l^{th}$ \textit{ConvBlock} is transformed correspondingly. For example, the scale weight is generated as follows: $
\gamma_{i,pri}^l{}' = Avg(w^l_{\cdot i}{}' \odot P^l_{\gamma}) = Avg(w^l_{\cdot p(i)} \odot P^l_{\gamma})= \gamma_{p(i),pri}^l. 
$

Similarly, we have $\beta_{i,pri}^l{}' =  \beta_{p(i),pri}^l$, $\mu_{i,pri}^l{}' =  \mu_{p(i),pri}^l$ and $\sigma_{i,pri}^l{}' = \sigma_{p(i),pri}^l$ in private passport of the $l^{th}$ \textit{ConvBlock}. As a result, $y^l_i{}' = y^l_{p(i)}$. However, the performance of $(l+1)^{th}$ \textit{ConvBlock} will be degraded as the learnable parameters are transformed with the wrong alignment unexpectedly, e.g.  $\gamma_{i,pri}^{l+1}{}'=Avg(\{w^{l+1}_{p(j),i} \}^{N_l}_{j=1} \odot P^{l+1}_{\gamma})$, which can not be calibrated by transforming other model parameters. 

For NeuronScale transformation on $l^{th}$ \textit{ConvBlock} in victim model, $\mathcal{NS}(B^l, \lambda)$ scales up/down the incoming/outgoing weights of each neuron to preserve the original model performance, i.e. $w^l_{c, i}{}' = \lambda_i w^l_{c, i}$ , $\lambda_i b^l_i = b^l_{i}$, and $w^{(l+1)}_{i,j}{}' = w^{(l+1)}_{i,j}/ \lambda_i$. Then, the parameter in private passport of the $l^{th}$ \textit{ConvBlock} is transformed correspondingly. For example, the scale weight is generated as follows: $\gamma_{i,pri}^l{}' = Avg(w^l_{\cdot i}{}' \odot P^l_{\gamma}) = Avg(\lambda_i w^l_{\cdot i} \odot P^l_{\gamma})= \lambda_i \gamma_{i,pri}^l. $

Similarly, we have $\beta_{i,pri}^l{}' =  \lambda_i \beta_{i,pri}^l$, $\lambda_i\mu_{i,pri}^l{}' =  \lambda_i \mu_{i,pri}^l$ and $\sigma_{i,pri}^l{}' = \lambda_i \sigma_{i, pri}^l$ in private passport of the $l^{th}$ \textit{ConvBlock}. As a result, $y^l_i{}' = \lambda_i y^l_{i}$. However, the performance of $(l+1)^{th}$ \textit{ConvBlock} will be degraded as the learnable parameters are transformed with the wrong alignment unexpectedly, e.g.  $\gamma_{i,pri}^{l+1}{}'=Avg(\{w^{l+1}_{j,i} / \lambda_j\}^{N_l}_{j=1} \odot P^{l+1}_{\gamma})$, which can not be calibrated by transforming other model parameters. 

\noindent\textbf{Passport-aware Normalization.}
This watermark scheme is applicable to batch normalization layer and generate the learnable parameters in private passport by the an additional FCN model with digital passports, i.e., $\gamma_{pri}^l = f(w^l \odot P^l_{\gamma}) $, $\beta_{pri}^l=f(w^l \odot P^l_{\beta})$, where $f$ is a fixed FCN provided by the owner in watermark verification stage. As our attack based on invariant transformations randomly converts the local feature of the convolution weights $w^l$, the parameters of $f$ need to be transformed correspondingly to align with the $w^l{}'\odot P^l_{\gamma}$ and $w^l{}'\odot P^l_{\beta}$ otherwise the result $\gamma_{pri}^l{}'$ and $\beta_{pri}^l$ can not be calibrated by modifying other weights in the victim model. However, it is impossible as the owner has no knowledge to the specific invariant transformations on the suspect model. As a result, the performance of the model with passport-aware branch drops quickly when we modify few neurons of the victim model which is illustrated in Fig.\ref{fig:passport_aware}.
\subsection{Implementation Details and More Evaluation Results} 
\label{sec:app:config}
\noindent\textbf{Hyper-parameters in Our Attack.} 
For each white-box watermark scheme evaluated in Section \ref{sec:case_study}, we follow the default settings in their source code released in \cite{code-aware,code-captioning,code-deepipr,code-deepsign,code-greedy,code-iprgan,code-lottery,code-riga,code-uchida} to reproduce the watermarked model. To obtain a LayerShuffle equivalence by $\mathcal{LS}(B,p)$, we randomly generate the permutation $p$ for each layer by the function $random.permutaion()$ in \textit{NumPy} \cite{numpy}. For NeuronScale operations $\mathcal{NS}(B,\lambda)$, we arbitrarily select the positive factor $\lambda_i$ from the candidate set $\{2^i\}^{16}_{i=0}$ to scale up/down the incoming/outgoing weights of each neuron. For layer-level SignFlip operation $\mathcal{SF}(B,s)$ used in Algorithm \ref{alg:attack frame}, we first generate an equal-length vector $x$ randomly sampled from the standard normal distribution and then obtain $s$ by extracting the signs of $x$. 

\noindent\textbf{Experimental Environments.}  All our experiments are conducted on a Linux server running Ubuntu 16.04, one AMD Ryzen Threadripper 2990WX 32-core processor and 1 NVIDIA GTX RTX2080 GPU.

\begin{figure}[h]
\centering
\subfigure
{
\begin{minipage}{\textwidth}
\centering
\includegraphics[width=0.85\textwidth]{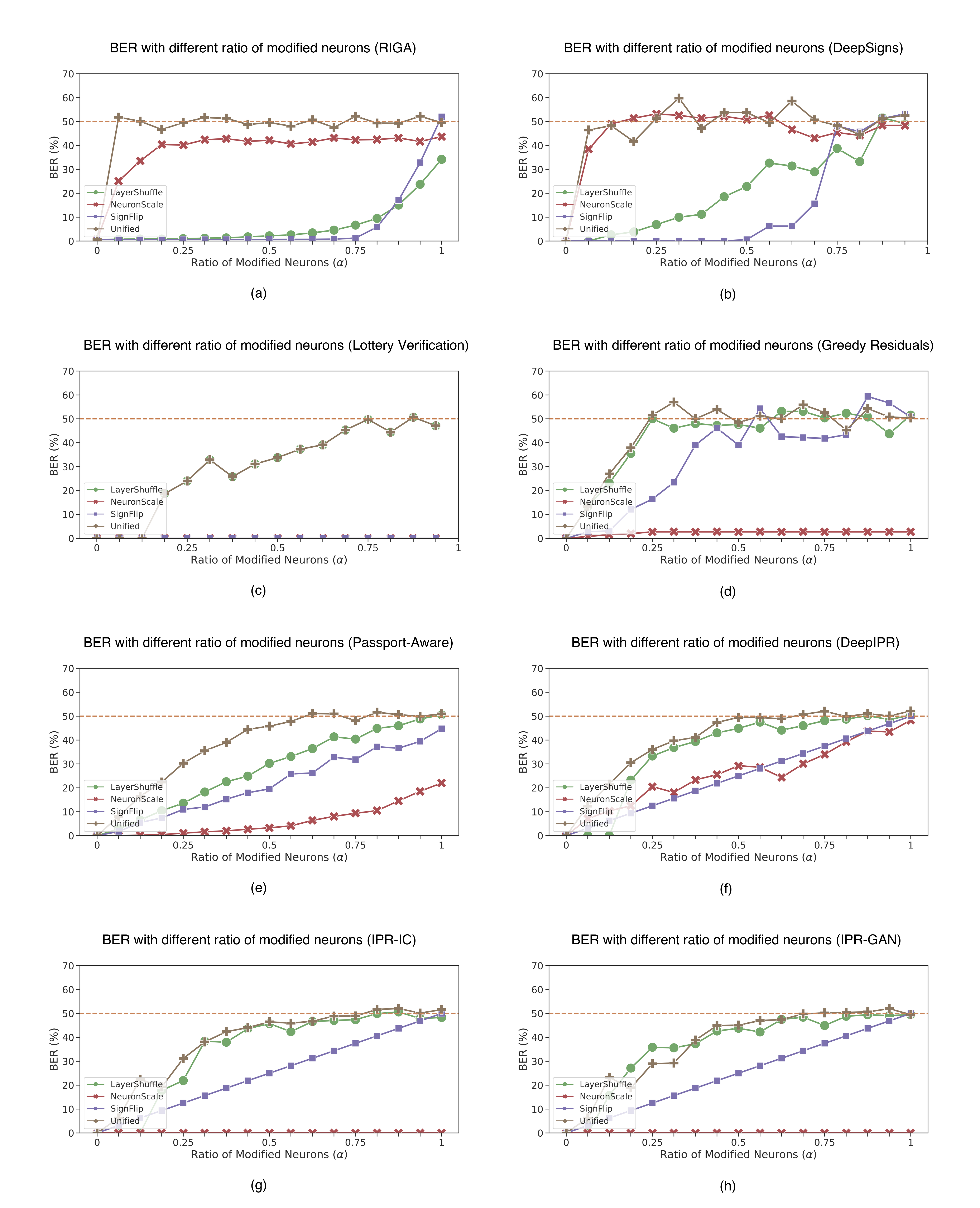}
\setlength{\abovecaptionskip}{0.cm}
\setlength{\belowcaptionskip}{-0.cm}
\caption{BER of the watermarked models after a certain proportion of neurons being modified by our attacks. The dashed horizontal lines reports the BER of the irrelevant model.
\label{fig:ber}}
\end{minipage}
}
\\
\centering
\subfigure
{
\begin{minipage}{\textwidth}
\centering
\includegraphics[width=0.4\textwidth]{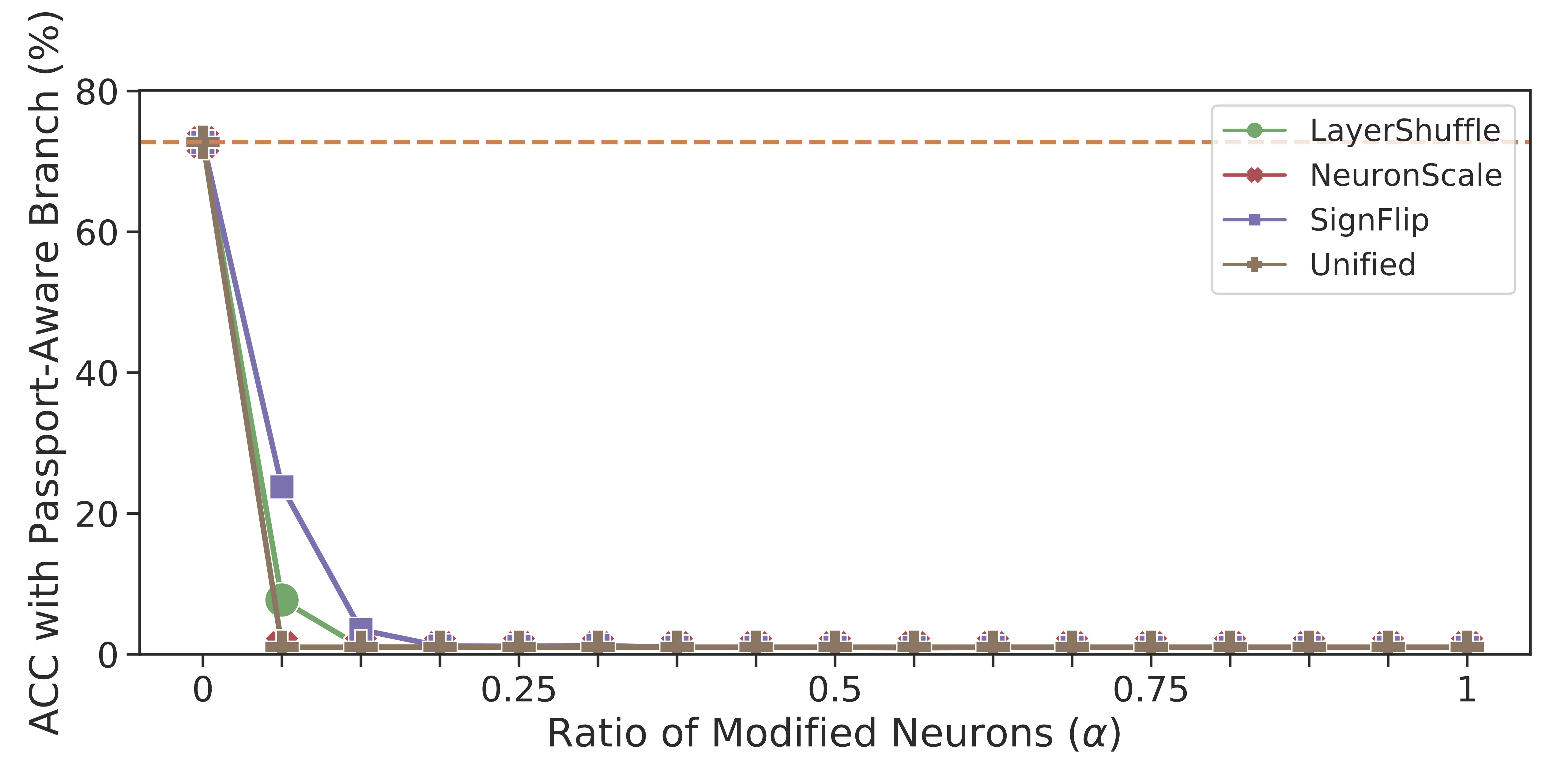}
\setlength{\abovecaptionskip}{0.cm}
\setlength{\belowcaptionskip}{-0.cm}
\caption{Performance of ResNet-18 (watermarked by Passport-aware Normalization) with the passport-aware branch when an $\alpha$ ratio of neurons are modified by our attacks, where the dashed horizontal line reports the original performance.}
\label{fig:passport_aware}
\end{minipage}
}
\end{figure}

\noindent\textbf{Omitted Results on Other Watermarking Schemes.}\label{sec:app:eval} 
In Fig.\ref{fig:ber}, we present the omitted results accompanying Fig.\ref{fig:ber_p_uchida} in the main text. In Fig.\ref{fig:passport_aware}, we present the omitted results accompanying Fig.\ref{fig:passport_deepipr} in the main text. 

\clearpage

\end{document}